\documentclass[preprint,aip,jcp]{revtex4-1}
\usepackage{epsfig}
\usepackage{textcomp}
\usepackage{graphicx}
\usepackage{amsmath}
\usepackage{amssymb}
\usepackage{amsmath}
\usepackage{url}
\usepackage[normalem]{ulem}

\usepackage{color}
\definecolor{Myblue}{rgb}{0.3,0.3,1.0}

\usepackage{xr}
\usepackage{rotating}

\begin{document}
\title{The ${\rm N}(^4S) +{\rm O}_2(X^3\Sigma^-_g) \leftrightarrow
  {\rm O}(^3P) + {\rm NO}(X^2\Pi)$ Reaction: Thermal and Vibrational
  Relaxation Rates for the $^{2}$A$'$, $^{4}$A$'$ and $^{2}$A$''$
  States}

\author{Juan Carlos San Vicente Veliz} \affiliation{Department of Chemistry,
  University of Basel, Klingelbergstrasse 80, CH-4056 Basel,
  Switzerland}
  
\author{Debasish Koner} \affiliation{Department of Chemistry,
  University of Basel, Klingelbergstrasse 80, CH-4056 Basel,
  Switzerland} 
  
\author{Max Schwilk} 
\affiliation{Department of Chemistry, University of Basel,
Klingelbergstrasse 80, CH-4056 Basel, Switzerland}

\author{Raymond J. Bemish} \affiliation{Air Force Research Laboratory,
  Space Vehicles Directorate, Kirtland AFB, New Mexico 87117, USA}

\author{Markus Meuwly} \email[]{m.meuwly@unibas.ch}
\affiliation{Department of Chemistry, University of Basel,
  Klingelbergstrasse 80, CH-4056 Basel, Switzerland}
\date{\today}

\begin{abstract}
The kinetics and vibrational relaxation of the ${\rm N}(^4S) +{\rm
  O}_2(X^3\Sigma^-_g) \leftrightarrow {\rm O}(^3P) + {\rm NO}(X^2\Pi)$
reaction is investigated over a wide temperature range based on
quasiclassical trajectory simulations on 3-dimensional potential
energy surfaces (PESs) for the lowest three electronic
states. Reference energies at the multi reference configuration
interaction level are represented as a reproducing kernel and the
topology of the PESs is rationalized by analyzing the CASSCF
wavefunction of the relevant states. The forward rate matches one
measurement at 1575 K and is somewhat lower than the high-temperature
measurement at 2880 K whereas for the reverse rate the computations
are in good agreement for temperatures between 3000 and 4100 K. The
temperature-dependent equilibrium rates are consistent with results
from JANAF and CEA results. Vibrational relaxation rates for O +
NO($\nu=1$) $\rightarrow$ O + NO($\nu=0$) are consistent with a wide
range of experiments. This process is dominated by the dynamics on the
$^2$A$'$ and $^4$A$'$ surfaces which both contribute similarly up to
temperatures $T \sim 3000$ K, and it is found that vibrationally
relaxing and non-relaxing trajectories probe different parts of the
potential energy surface. The total cross section depending on the
final vibrational state monotonically decreases which is consistent
with early experiments and previous simulations but at variance with
other recent experiments which reported an oscillatory cross section.
\end{abstract}
 
\maketitle

\section{Introduction}
Reactions involving nitrogen and oxygen play important roles in
combustion, supersonic expansions, hypersonics, and in atmospheric
processes. A particularly relevant process, which is part of the
so-called Zeldovich process\cite{zeldovich:1946} are the NO + O or
O$_{2}$ + N reactions\cite{Bose1997,Dodd1999} that describe the
oxidation of nitrogen. In the forward direction, the reaction also
generates reactive atomic oxygen. These reactions, together with a
range of other atom plus diatom and diatom plus diatom reactions form
the core of the 5- and 11-species model used in
hypersonics.\cite{gupta:1990} At high temperatures ($\sim 20000$ K),
as present in thin regions of shock layers created at hypersonic speed
flight\cite{park1993review}, the reactive chemical processes can
become very complex. This complexity is in part due to a significant
degree of non-equilibrium. The lack of experimental information on the
kinetics at these high temperatures makes numerical simulations for
reaction cross sections as well as reaction and vibrational relaxation
rates a very valuable source of information for characterizing
hypersonic flow.\\

\noindent
There is also much interest in correctly describing the vibrational
distribution of the NO molecules after reactive or nonreactive
collisions with atomic oxygen for atmospheric processes. The infrared
emission of nitric oxide is one of the main tracers to follow and
characterize the energy budget in the upper
atmosphere.\cite{bailey:2016} This emission arises from relaxation of
vibrationally excited NO after collisional excitation with atomic
oxygen. This relaxation process has also been implicated in nighttime
cooling of the thermosphere, above $\sim 100$ km.  Furthermore, nitric
oxide is also formed in situ and used as a tracer for combustion and
in hypersonic flows where it is commonly observed by Laser Induced
Fluorescence (LIF).  \\

\noindent
Previous studies included experimental and computational
characterizations of the reaction dynamics and final state
distributions of the products. Using a pulsed beam of energetic
nitrogen atoms at 8 km/s interacting with thermal oxygen under single
collision conditions to mimic velocities seen in low earth orbit , the
distribution of vibrationally excited NO and state specific reaction
cross sections for N+O$_2$ $\rightarrow$ NO+O were
determined.\cite{Caledonia:2000} The analysis showed an oscillatory
behaviour of the cross section with increasing final vibrational
state, with minima at $\nu=3$ and $\nu=6$, with an uncertainty of a
factor of two. An even earlier experiment\cite{Winkler1986}, using
saturated multiphoton ionization spectroscopy, measured the NO product
ground-state distribution, reporting a difference in the cross
sections between odd ($\nu = 1,3,5$) and even ($\nu = 0,2,4,6$) final
vibrational levels. \\

\noindent
From the perspective of computer based
simulations\cite{Caridade2004,Duff1994,Ramachandran2000} the
vibrational state-dependent cross sections have been calculated
using a variety of potential energy surfaces (PESs). In all of these
computational studies, the maximum of the final state vibrational
cross section is found to be at $\nu = 1$\cite{Caridade2004} or at
$\nu=2$\cite{Duff1994,Ramachandran2000} with no notable
oscillation. One PES for the $^2$A$'$ state used a fit\cite{Duff1994}
to electronic structure calculations at the complete active space SCF
(CASSCF) level followed by multireference contracted configuration
interaction and a modified Duijneveldt (11s6p) basis
set.\cite{walch:1987} Another PES was based on 1250 (for the $^2$A$'$)
and 910 ($^4$A$'$) CASPT2 calculations and fitted to an analytical
function.\cite{Says2002} Such an approach was also used for the
$^2$A$''$ state.\cite{Gonzlez2001} This was followed by a PESs for the
$^2$A$'$ state using a diatomics in molecules (DIM) expansion with the
two-body terms based on extended Hartree-Fock
calculations.\cite{varandas:2003} Then, a 2-dimensional PES with the
NO bond length fixed at its equilibrium value of 2.176 a$_0$ was
determined at the icMRCI+Q level of theory and a cc-pVQZ and
represented as a cubic spline.\cite{Ivanov2007} This work also
presented a PES for the $^2$A$''$ state. More recently, a double many
body expansion fit to 1700 points at the MRCI/aug-cc-VQZ level of
theory for the $^2$A$''$ state was carried out.\cite{Mota2012} In
addition, quasi classical trajectory (QCT)
calculations\cite{Caridade2004,CastroPalacio2014,Bose1997,Says2002}
have been reported for the temperature dependent rate for the
N($^4$S)+O$_{2}$ $\rightarrow$ NO+O and its reverse reaction using
different PESs. \\

\noindent
Another important process is the energy transfer following the
collision of vibrationally excited NO with oxygen atoms (O$_{\rm A}$ +
NO$_{\rm B}$ $\rightarrow$ O$_{\rm A}$ + NO$_{\rm B}$) or (O$_{\rm A}$
+ NO$_{\rm B}$ $\rightarrow$ O$_{\rm B}$ + NO$_{\rm A}$) to yield NO
in its ground vibrational state. Using 355 nm laser photolysis of a
dilute mixture of NO$_2$ in argon, the experiment\cite{Dodd1999}
reports a vibrational relaxation rate of: $k_{\nu=1 \rightarrow 0}=
2.4 \pm 0.5 (10^{-11})$ cm$^{3}$s$^{-1}$ at a temperature of $T=298$
K. Later, QCT simulations\cite{Caridade2008} reported a value of
$k_{\nu=1 \rightarrow 0} = 2.124 \pm 0.73 (10^{-11})$ cm$^{3}$s$^{-1}$
at $T=298$ K which is close to the experimentally reported
rate. Another experiment\cite{Hwang2003} used a continuous wave
microwave source to form O atoms combined with photolysis of trace
amounts of added NO$_{2}$ to generate vibrationally excited NO. This
experiment found a rate of $k_{\nu=1 \rightarrow 0} = 4.2 \pm 0.7
(10^{-11})$ cm$^{3}$s$^{-1}$ at $T=295$ K which is larger by 75 \%
compared with the earlier experiments.\cite{Dodd1999} Quite recent QCT
simulations using again the DIM-based PES\cite{varandas:2003}
mentioned above reported a rate of $k_{\nu=1 \rightarrow 0} = 4.34 \pm
0.7 (10^{-11})$ cm$^{3}$s$^{-1}$ at $T=298$ K from QCT
simulations\cite{Caridade2018} which used the empirical DIM PES for
the $^2$A$'$ ground state\cite{varandas:2003} and a more recent,
MRCI-based fitted PES for the $^2$A$''$ state.\cite{Mota2012}\\

\noindent
Given the rather heterogeneous situation for the quality of the
existing PESs for studying the ${\rm N}(^4S) +{\rm O}_2(X^3\Sigma^-_g)
$ reaction and the vibrational relaxation of NO, the present work
determines fully dimensional PESs using a consistent methodology to
represent the 3 lowest electronic states, $^{2,4}$A$'$ and $^{2}$A$''$
as well as to evaluate the cross sections for the (forward)
\begin{equation}
{\rm N}(^4S) +{\rm O}_2(X^3\Sigma^-_g) \rightarrow {\rm O}(^3P) + {\rm
  NO}(X^2\Pi)
\end{equation}
and (reverse)
\begin{equation}
{\rm O} (^3P) + {\rm NO} (X^2 \Pi) \rightarrow {\rm N} (^4S) + {\rm
  O}_2 (X^3 \Sigma _g^-)
\end{equation}
reaction. All three states are energetically accessible in the
hypersonic regime, i.e. at temperatures up to 20000 K. Experimentally, cross sections and rates
for the forward and reverse reactions have been measured and
experimental data for vibrational relaxation rates are
available\cite{Caledonia:2000,Winkler1986,Dodd1999,Hwang2003,CEA1996,Chase1985}
which serve as benchmarks for the present work.\\

\noindent
In the following, the calculation and representation of the asymptotic
PESs for the three electronic states and the two channels are
described. These are combined to a set of fully-dimensional, reactive
PESs which are suitable for quasiclassical trajectory simulations from
which cross sections, reaction rates and rates for vibrational
relaxation can be determined.  The results of the simulations are
discussed in the context of the limits of errors in the simulations
and comparisons with available experimental data. Finally, the basis
of the observables is discussed at an atomistic level, based on
analyzing the trajectories.\\
 
\section{Computational Methods}
\noindent
This section presents the generation and representation of the
potential energy surfaces and the methodologies for the quasiclassical
trajectory (QCT) simulations and their analysis. All PESs are computed
at the multi reference CI (MRCI) level of theory together with large
basis sets. These are then exactly represented using the reproducing
kernel Hilbert space approach. The quality of the representation is
then checked using additional MRCI calculated points.\\

\subsection{The $^2$A$'$, $^2$A$''$ and $^4$A$'$ Potential Energy Surfaces}
{\it Ab initio} energy calculations were carried out for the
$^{2}$A$'$, $^2$A$''$ and $^{4}$A$'$ states. The energies were
computed on a grid defined by Jacobi coordinates $(r,R,\theta)$ where
$r$ is the separation of the diatomic, $R$ is the distance between the
atom and the center of mass of the diatomic and $\theta$ is the angle
between the two unit vectors $\vec{r}$ and $\vec{R}$. For $R$ the grid
included 28 points between 1.4 and 12.0 a$_{0}$, the distance $r$ was
covered by 20 points between 1.5 and 4.0 a$_{0}$ and the angular grid
contained 13 angles from a Gauss-Legendre quadrature (169.796,
156.577, 143.281, 129.967, 116.647, 103.324, 90.100, 76.676, 63.353,
50.033, 36.719,23.423, 10.204). In order to consistently describe all
relevant states and avoid numerical instabilities due to closely-lying
states of the same symmetry, state-averaged CASSCF\cite{wen85:5053,kno85:259,wer80:2342} calculations
including the two lowest states of each symmetry (two spin symmetries
and two spatial symmetries) were carried out. Hence, in total eight
states are included in the CASSCF reference wave function. A
subsequent MRCISD\cite{wer88:5803,kno88:514} (referred to as MRCI+Q in the following) calculation
of the lowest state for each symmetry then computes dynamical electron
correlation contributions at a high order level. The augmented Dunning-type
correlation consistent polarize triple zeta (aug-cc-pVTZ)\cite{dun89:1007}
basis set is used in this work. All electronic structure calculations are
done with the Molpro-2018 \cite{MOLPRO_brief} software package.
For each of the electronic states, {\it ab initio} energy calculations
have been performed for total 7280 points for the NO+O channel and
3920 (including symmetry) points for the OO+N channel, i.e. overall $\sim
11000$ points which is more than 5 times more reference calculations
compared with previous efforts at a similar level of theory. It is to
be noted that electronic structure calculations for a fraction of the
geometries at large $R$ and/or $r$ converged to excited
states. Those points were excluded from the training energy data
set.\\

\noindent
For certain geometries ($<$ 0.5 \%)  outside the equilibrium region
the CASSCF or MRCI calculations did not
converge. In these cases, the missing grid points were
reconstructed using a 2-dimensional reproducing kernel ($R,r$)
(RKHS)\cite{Unke2017} for each $\theta$. This procedure of
discriminating possible outliers was necessary before constructing the
full dimensional PES. The 3-dimensional PES for each channel
$V(R,r,\theta)$, is constructed using a reciprocal power decay kernel
with $n= 2$ and $m= 6$ for the two radial coordinates and an Taylor
spline kernel with $n= 2$ for the angular part.\cite{Unke2017} The
regularization parameter used was $\lambda = 10^{-18}$.\\

\noindent
The global, reactive 3D PES $V(r_1,r_2,r_3)$ for an electronic state is constructed
by summing the individual PESs for each channel 
\begin{equation}
V(r_1,r_2,r_3)=\sum_{j=1}^{3} w_{j} (r_{j}) V_{j} (R,r_j,\theta),
\label{eq:mixed}
\end{equation}
using an exponential switching function with weights
\begin{equation}
w_{i} (r)=\frac{e^{-(r_{i}/dr_{i})^{2}}}{\sum_{j=1}^{3}
  e^{-(r_{j}/dr_{j})^{2}}}.
\end{equation}
Here, $dr_{i}$ are switching function parameters for the two channels
(I) O$_2$ + N and (II) NO + O.  These parameters were optimized by a
least square fit to obtain values of (1.25, 1.11, 1.11) a$_0$, (1.07,
0.87, 0.87) a$_0$ and (1.40, 1.35, 1.35) a$_0$ for the $^2$A$'$,
$^4$A$'$ and $^2$A$''$ PESs, respectively.\\

\noindent
The global, local minima and transition states between the minima
and/or entrance channels supported by the PESs were determined using
BFGS minimization and the nudged elastic band
method\cite{jonsson:2000} as implemented in the atomic simulation
environment (ASE).\cite{Larsen:2017}\\

\subsection{Quasi-Classical Trajectory Simulations}
The QCT simulations used in the present work have been extensively
described in the literature\cite{tru79,hen11,kon16:4731,Koner2018}.
Here, Hamilton's equations of motion are solved using a fourth-order
Runge-Kutta numerical method. The time step was $\Delta t = 0.05$ fs
which guarantees conservation of the total energy and angular
momentum. Initial conditions for the trajectories are sampled using
standard Monte Carlo sampling method.\cite{tru79} The reactant and
product ro-vibrational states are determined following semiclassical
quantization. Since the ro-vibrational states of the product diatom
are continuous numbers, the states are assigned by rounding to integer
values. Two schemes were used 1) histogram binning (HB), i.e. rounding
values to the nearest integers, or 2) Gaussian binning (GB), which
weights each trajectory with a Gaussian shaped function (with a full
width at half maximum of 0.1) centered on the integer
values.\cite{bon97:183,bon04:106,kon16:4731} Here, both schemes were
tested and found to yield comparable results. Therefore results
obtained from GB are reported in the following.\\

\noindent
The state-to-state reaction cross section at fixed collision energy
$E_{\rm c}$ is $\sigma_{v,j \rightarrow v',j'}(E_{\rm c}) = 2 \pi
\int_{0}^{b_{\rm max}} P_{v,j \rightarrow v',j'}(b;E_c) b db$. This
integral can be evaluated using Monte Carlo sampling\cite{tru79} which
yields
\begin{equation}
\sigma_{v,j \rightarrow v',j'}(E_{\rm c}) = \pi b^2_{\rm max}
\frac{N_{v',j'}}{N_{\rm tot}},
\end{equation}
where $N_{v',j'}$ is the number of reactive trajectories corresponding
to the final state $(v',j')$ of interest, $N_{\rm tot}$ is the total
number of trajectories, $P_{v,j \rightarrow v',j'} = N_{v',j'}/N_{\rm
  tot}$ is the probability to observe a particular transition $(v,j)
\rightarrow (v',j')$, and $b_{\rm max}$ is the maximum impact
parameter for which a reactive collision occurs. Here, $b_{\rm max}$
is calculated by running batches of trajectories at different
intervals of $b$. In the present work stratified
sampling\cite{tru79,ben15:054304} is used to sample the impact
parameter $b \in [0 \leq b \leq b_{\rm max}]$. The sampling strategy
is described in detail in previous work.\cite{Koner2018}\\

\noindent
The thermal rate for an electronic state ($i$) at a given temperature ($T$) is then obtained from
\begin{equation}
 k_i(T) = g_i(T)\sqrt{\frac{8k_{\rm B}T}{\pi\mu}} \pi b^2_{\rm max}
 \frac{N_{r}}{N_{\rm tot}},
 \label{eq8}
\end{equation}
where $g_i(T)$ is the electronic degeneracy factor of electronic state `$i$', $\mu$ is the reduced
mass of the collision system, $k_{\rm B}$ is the Boltzmann constant,
and, depending on the specific process considered, $N_r$ is the number
of reactive or vibrationally relaxed trajectories. In the rate
coefficient calculations, the initial ro-vibrational states and
relative translational energy ($E_{\rm c}$) of the reactants for the
trajectories are sampled from Boltzmann and Maxwell-Boltzmann
distribution at a given $T$, respectively. The sampling methodology is
discussed in detail in Ref. \cite{Koner2018}.\\

\noindent
For the forward reaction (N($^{4}$S) + O$_2$(X$^3 \Sigma^-_g$)
$\rightarrow$ O($^3$P) +NO(X$^2 \Pi$)) the rate $k_{+}(T)$
is calculated using degeneracies of 1/6 and 1/3 for the $^2$A$'$ and $^4$A$'$ states,
respectively, whereas for the reverse reaction (O($^{3}$P) + NO(X$^{2}
\Pi$) $\rightarrow$ N($^4$S) +O$_2$(X$^3 \Sigma _g^-$)) the degeneracies are
\begin{equation}
g_{^2A'}(T)=\frac{2}{(5+3 \cdot
  e^{\frac{-227.8}{T}}+e^{\frac{-326.6}{T}})(2+2e^{\frac{-177.1}{T}})}
  \label{ge1}
\end{equation}
and
\begin{equation}
g_{^4A'}(T)=\frac{4}{(5+3 \cdot
  e^{\frac{-227.8}{T}}+e^{\frac{-326.6}{T}})(2+2e^{\frac{-177.1}{T}})}
  \label{ge2}
\end{equation}
The terms in Eqs. \ref{ge1} and \ref{ge2} are the
degeneracies of the $J$ or spin states and the exponential parameters
227.8, 326.6 and 177.1 are the energy differences (in units of K) between two
neighboring states. The equilibrium constant is then
\begin{equation}
K_{eq}(T)=\frac{k_{+}(T)}{k_{-}(T)}.
\label{eq:equilibrium}
\end{equation}

\section{Results}
\subsection{The Potential Energy Surfaces}
\begin{figure}[h!]
\centering \includegraphics[scale=0.8]{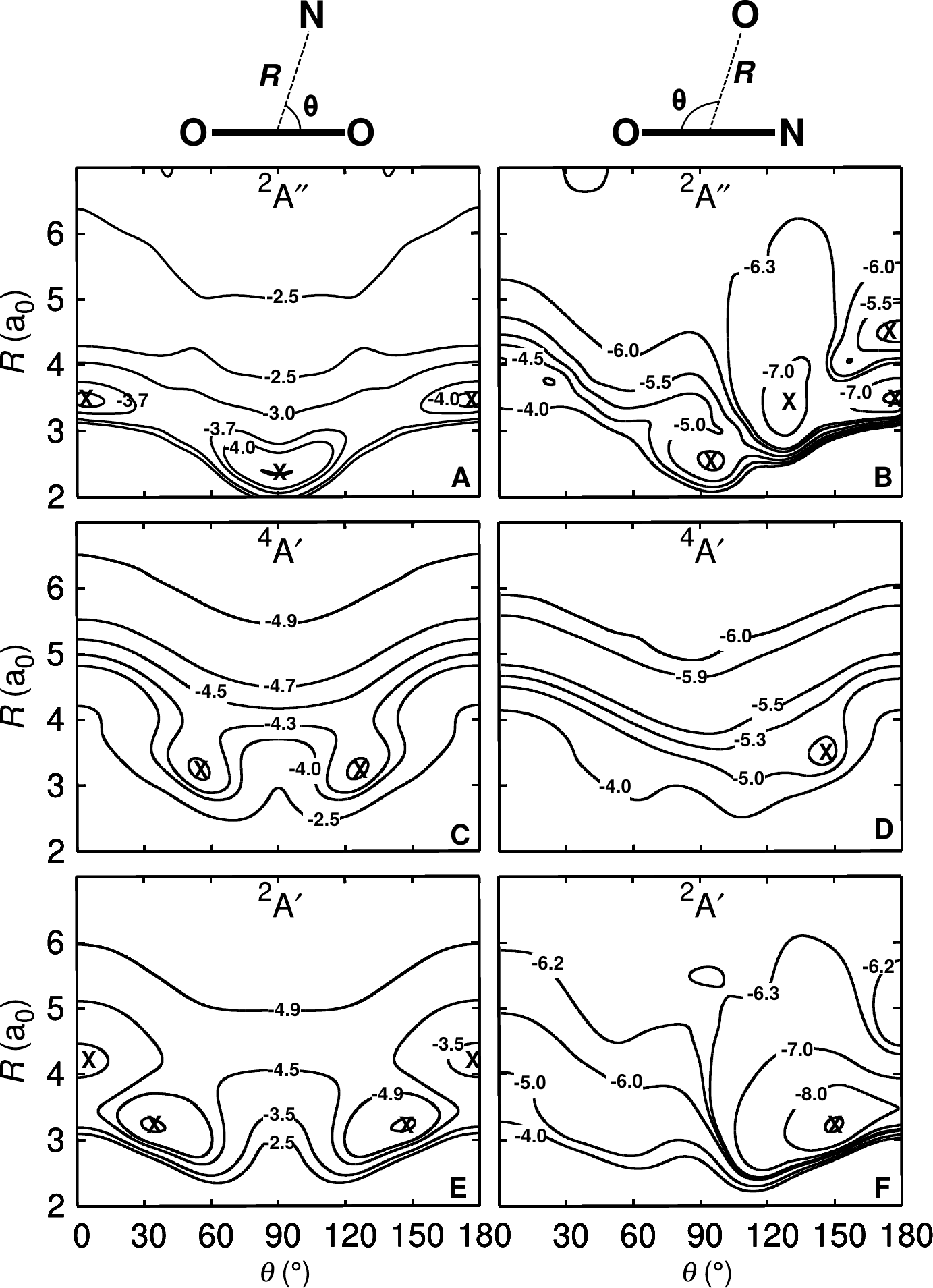}
\caption{Two-dimensional cuts through the 3-d PES for the OO+N (left
  and the NO+O (right) channels. The OO and NO diatomics are at their
  equilibrium bond lengths of the respective states, see Figures
  S2 to S4. They are $R^{\rm (OO)} = 2.33$
  \AA\/, $R^{\rm (OO)} = 2.39$ \AA\/, $R^{\rm (OO)} = 2.30$ \AA\/, and
  $R^{\rm (NO)} = 2.26$ \AA\/, $R^{\rm (NO)} = 2.36$ \AA\/, $R^{\rm
    (NO)} = 2.28$ \AA\/ for the $^2$A$'$, $^4$A$'$, and $^2$A$''$
  states, respectively, from bottom to top. Specific contours with
  energies in eV are indicated. The zero of energy is for dissociation
  into atomic fragments O$(^3{\rm P})$+O$(^3{\rm P})$+N$(^4{\rm
    S})$. The symbols indicate the minima discussed in the text and
  the definition of the coordinates is given on top of the Figure. For
  the NO+O asymptote the OON geometry corresponds to $\theta = 0$
  whereas ONO has $\theta = 180^\circ$.}
\label{fig:pes}
\end{figure}

An overview of the PESs, see Figures \ref{fig:pes}, S1,
and Table \ref{tab:params}, for all three states investigated
($^{2}$A$'$, $^{4}$A$'$, and $^{2}$A$''$ from bottom to top) is given
as 2-dimensional projections with the two asymptotes (N+OO and O+NO)
on the left and right columns in Figure \ref{fig:pes},
respectively. It should be noted that these representations are all
for diatomic separations (O$_2$ and NO, respectively) at values of
critical points (see Figures S2 to S4) and
therefore do not exhibit all features of the full 3-dimensional PES.\\

\noindent
All PESs for OO+N are symmetric with respect to $\theta = 90^\circ$,
as expected. For the $^2$A$'$ state and the O$_2$+N dissociation limit
the 2d-PES was generated for TS1 in Figure S2, i.e. for
$R^{\rm (OO)} = 2.33$ a$_0$. The two symmetry related minima are at
$R_{e}^{\rm (NO)} = 3.23$ a$_0$ and $\theta = 34^\circ$ and $\theta =
146^\circ$, respectively. For the NO+O dissociation limit the PES with
$R^{\rm (NO)} = 2.26$ a$_0$ (corresponding to MIN3 in Figure
S2) displays two minima. They are at $(R^{\rm (OO)} =
3.38$ a$_0$, $\theta = 35^\circ)$ and $(R^{\rm (OO)} = 3.22$ a$_0$,
$\theta = 150^\circ$).\\

\noindent
For the $^4$A$'$ state the surface for the O$_2$+N dissociation limit
has $R^{\rm (OO)} = 2.39$ a$_0$ for TS1 in Figure S3 and
the 2-dimensional PES in Figure \ref{fig:pes} has the minimum at
$R^{\rm (NO)} = 3.19$ a$_0$ with $\theta = 56^\circ$ and $\theta =
124^\circ$, respectively. Conversely, at the NO+O asymptote, the PES
is almost purely repulsive for $R^{\rm (NO)} = 2.36$ a$_0$ (TS2 in
Figure S3). In Jacobi coordinates, a faint minimum is at
($R^{\rm (OO)} = 3.48$ a$_0$, $\theta = 146^\circ$).\\

\noindent
Finally, for the $^2$A$''$ state the 2-dimensional PES is reported for
$R^{\rm (OO)} = 2.30$ a$_0$ in the OO+N channel (TS1 in Figure
S4). It exhibits two minima at ($R^{\rm (NO)} = 2.36$
a$_0$, $\theta = 90^\circ$), and ($R^{\rm (NO)} = 3.55$ a$_0$, $\theta
= (0,180)^\circ$). At the NO+O asymptote the PES has multiple minima,
see Figure \ref{fig:pes}. For $R^{\rm (NO)} = 2.28$ a$_0$ they are at
($R^{\rm (OO)} = 3.69$ a$_0$, $\theta=0^\circ$), ($R^{\rm (OO)} =
2.53$ a$_0$, $\theta=95^\circ$), ($R^{\rm (OO)} = 3.28$ a$_0$,
$\theta=128^\circ$), and ($R^{\rm (OO)} = 3.50$ a$_0$, $\theta =
180^\circ$).\\

\begin{tiny}
\begin{table}[ht!]
  \caption{Minima (MIN$i$) and transition states (TS$i$) were
    calculated using the Nudged Elastic Band
    (NEB)\cite{jonsson:2000,hammer:2016} method. Equilibrium distances
    in a$_{0}$, angle in degree for $\angle ({\rm NOO})$ and $^{a}
    \angle ({\rm ONO})$, and energies $\Delta E_1$ (in eV) with
    respect to the N+O+O asymptote and $\Delta E_2$ (in kcal/mol)
    relative to the N + O$_{2}$ limit, except for $^{b}$(with respect
    to the global minimum), and $^{c}$(relative to the O+NO limit) to
    compare with the literature. For the energy level diagram and the
    connectivities, see Figures S2 to S4.}
    \begin{tabular}{l|rrrr||r|r}
    \hline
    \hline
    $^{2}$A$'$ & $R_{e}^{\rm (NO)}$ & $R_{e}^{\rm OO}$ & $<$NOO & $\Delta E_1$  & $\Delta E_2$ & $\Delta E_2$\cite{Says2002}\\
    \hline
    MIN1 & 2.27 & 2.58 & 130.4  &  --5.78 & --18.34 & --28.50  \\
    TS1  & 3.46 & 2.33 & 112.1  &  --4.64 & 8.07  &    6.87 \\
    TS2  & 2.20 & 2.86 & 134.1  &  --5.71 &--16.82 & --27.42  \\
    TS3  & 2.19 & 4.63 &  90.7  &  --6.29 &--30.12 & --34.26 \\
    MIN2 & 2.18 & 4.51 & 121.1  &  --6.33 &--31.10 & --37.64 \\
    MIN3 & 2.26 & 4.17 &  22.9  &  --9.46 &--103.20 & --108.68 \\
    \hline
 $^{4}$A$'$ &  &  &  &   &  & $\Delta E_2$\cite{Says2002}\\
    \hline
    MIN1 & 2.65 & 2.60 & 104.0 & --4.69 & 6.67 &  5.43 \\
    TS1   & 3.26 & 2.39 & 108.0 & --4.33 & 15.01 & 12.74  \\
    TS2   & 2.36 & 3.07 & 103.0 & --4.68 & 8.75 &  7.81  \\
    \hline
       $^{2}$A$''$ &  &  &  &  &  & $\Delta E_2$\cite{Mota2012} \\
    \hline
     MIN2 & 2.23 & 4.06 & 107.5$^{a}$ & --7.37 & --111.79 & --113.95  \\
     TS1  & 3.96 & 2.30 & 111.7       & --2.48 &  0.93 & 0.63  \\
     TS2  & 4.35 & 2.31 & 30.8$^{a}$ & --2.44  &   1.95 & 2.07  \\
     TS3  & 2.57 & 4.35 & 80.2$^{a}$ & --5.54  & -69.61  & --77.81\\
     TS4  & 2.42 & 3.99 & 109.4$^{a}$ & --7.22 & --108.34  & --113.14\\
     TS5  & 2.39 & 4.31 & 130.2$^{a}$ & --6.78 & 18.60$^{b}$  & 32.25$^{b}$\\
     MIN1 & 2.61 & 2.88 &  67.1$^{a}$ & --6.16 &--83.93 & --85.59 \\
     MIN3 & 2.28 & 4.56 &  180.0$^{a}$ & --7.58 &--29.06$^{c}$ & --38.41$^{c}$ \\
    \hline
    \hline
  \end{tabular}
  \label{tab:params}
\end{table}
\end{tiny}

\noindent
All minima and transition states together with their connectivities on
the 3d PES are given in the supplementary information, see Figures
S2 to S4. Several paths which include a
number of minima and transition states can be found on the $^2$A$'$
and $^2$A$''$ PESs for the forward and reverse reaction while both
reactions follow rather simple paths on the $^4$A$'$ PES. It is
worthwhile to note that there are no crossings between the $^2$A$'$
and $^2$A$''$ PESs as well as between $^2$A$'$ and $^4$A$'$ electronic
states which differs from, e.g., the [CNO]-system.\cite{Koner2018}\\

\begin{figure}[h!]
\centering
\includegraphics[scale=0.7]{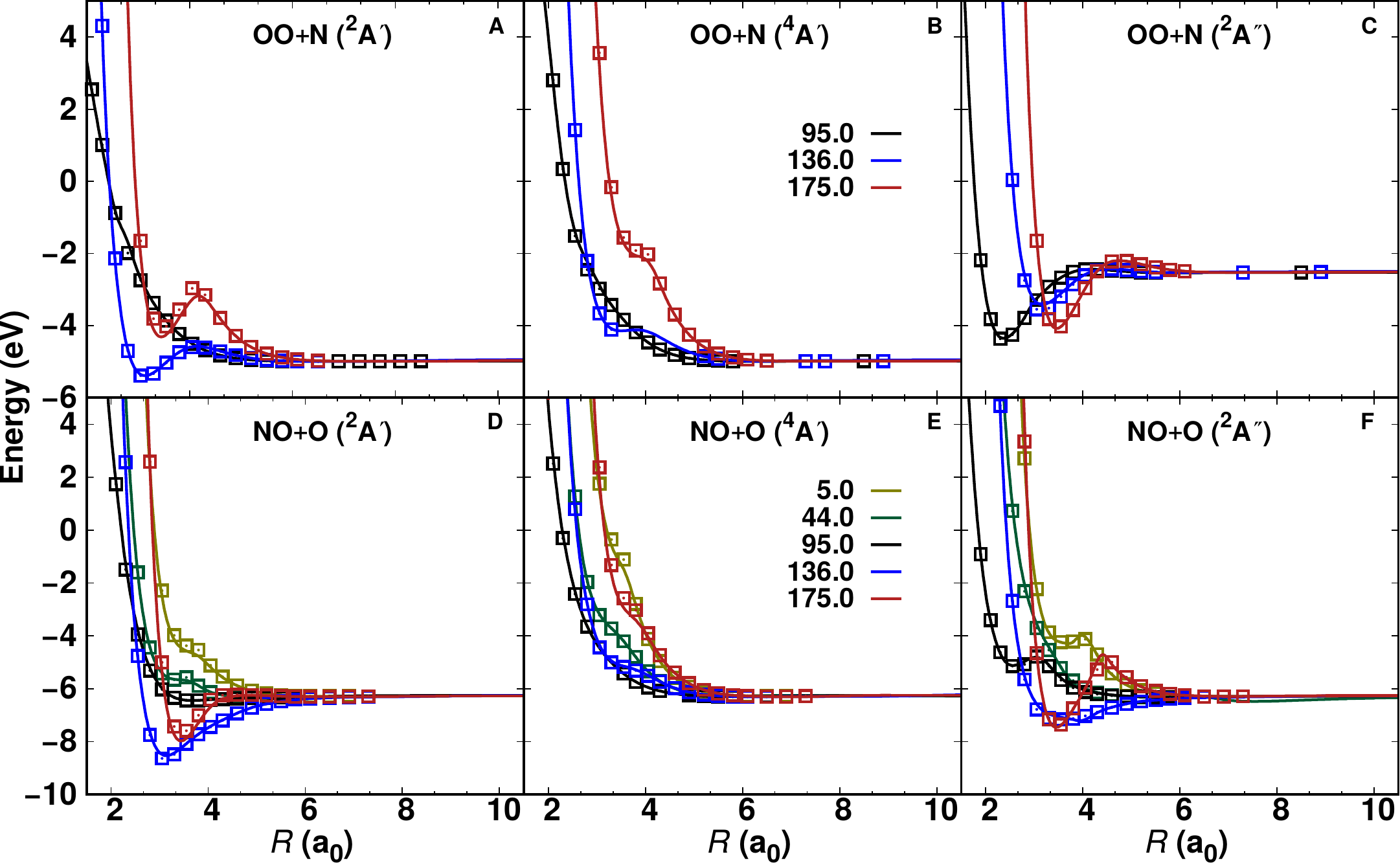}
\caption{Quality of the RKHS representation of the 3d PESs at off-grid
  points. The MRCI+Q/aug-cc-pVTZ reference energies (open symbols) and
  the RKHS interpolated energies (lines) for the $^{2}$A$'$,
  $^{4}$A$'$ and $^{2}$A$''$ for the OO+N (top, $r^{\rm (OO)} = 2.30$
  a$_0$) and NO+O (bottom, $r^{\rm (NO)} = 2.19$ a$_0$) channels are
  reported.}
\label{fig:1dpes}
\end{figure}

\noindent
One-dimensional cuts along the O$_2$+N and NO+O coordinates for
constant angle $\theta$ for the three different electronic states are
reported in Figure \ref{fig:1dpes}. All angular cuts correspond to
off-grid points, i.e. data not explicitly used in generating the
RKHS. Therefore, the RKHS energies (solid lines) are predictions and
are found to compare well with the true energies calculated at the
MRCI+Q/aug-cc-pVTZ level of theory. Nevertheless, for a few points on
the $\theta = 175.0^\circ$ cut around $R \sim 4$ a$_0$ for the $^2$A$'$
state (see Figure \ref{fig:1dpes}A) the RKHS-predicted energies differ
slightly from the true energies.\\

\noindent
The quality of all three PESs for both, on- and off-grid points is
reported in Figure S5 as correlation plots. The
correlation between the reference ({\it ab initio} energies) and RKHS
energies ranges from $R^2 = 0.9996$ to 0.9999 for grid points and from
$R^2 = 0.9992$ to 0.9997 for off-grid points for the three electronic
states. The corresponding root mean squared errors for the on-grid
points range from 0.022 to 0.043 eV and off-grid points from 0.033 to
0.057 eV.  It should be noted that all the RKHS energies are evaluated
on the mixed, fully reactive PES, see Eq. \ref{eq:mixed}. The
agreement between reference and RKHS energies is even better if the
channels are considered separately. \\

\begin{figure}[h!]
\centering
\includegraphics[scale=0.55]{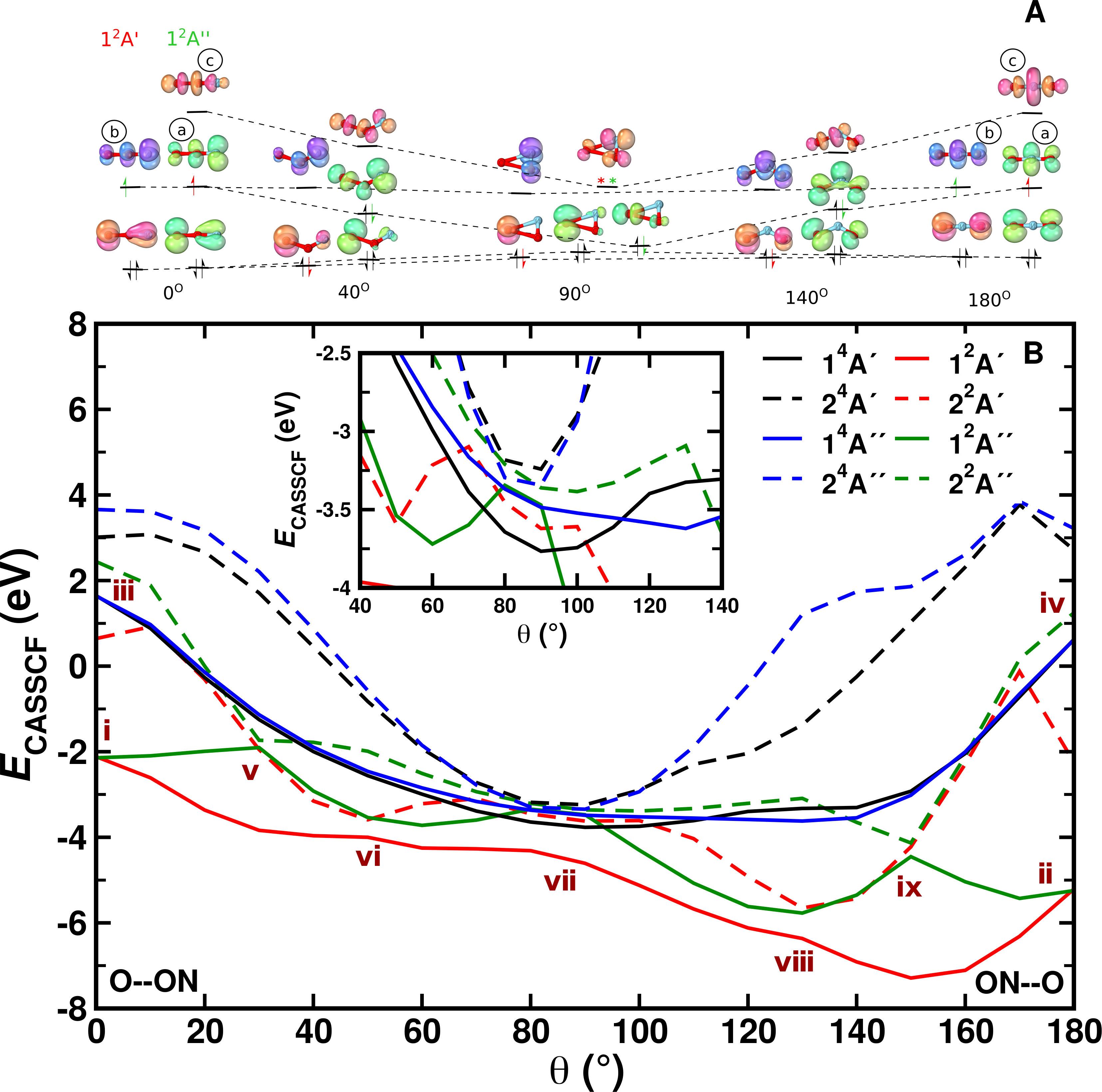}
\caption{Panel A: MO diagram of NO$_2$ for the doublet ground state
  for varying values of $\theta$. The dominant configurations at
  selected angles are shown for the lowest $^2$A$'$ and $^2$A$''$
  states (in black occupations occurring for both states,
  state-specific orbital occupations are colour-coded). An asterisk
  indicates significant (additional) occupation of an orbital due to
  strong electron correlation. Details for each of the states are
  provided in Figure S6. Panel B: Energies in eV
  relative to separated atoms. The inset shows details of the states
  around the T-shaped geometry. Features i through ix are discussed in
  the text.}
\label{fig:mo}
\end{figure}

\noindent
To rationalize the observed topology of the NO+O channel of the MRCI+Q
PES (Figure \ref{fig:pes} panels B, D, and F), the orbital diagram of
the natural orbitals as obtained from the CASSCF calculations are
analyzed. Figure \ref{fig:mo} shows the evolution of the natural
orbitals and the energies for $R=3.4$ a$_0$ and $r_{\rm NO} = 2.183$
a$_0$ (equilibrium NO separation) with varying values of
$\theta$. Only natural orbitals with significant change in occupation
number are shown along the path. Figure S7 shows a
complete MO diagram of the valence space. The dominant configurations
for the lowest $^2$A$'$ and $^2$A$''$ states are indicated in the MO
diagrams, and an illustration of all main configurations along the
path is given in Figure S6. The cut qualitatively
includes most of the stationary states of the 2-dimensional PESs of
the NO+O channel (see Figures \ref{fig:pes} and S5).
For the linear structures (Figure \ref{fig:mo}A) two perpendicular
$\pi_3$-systems arise with one electron in an antibonding $\pi_3$*
orbital. The bonding orbital of the $\pi$-system shows a more equal
contribution from all three atomic centers for the linear ONO
structure than for the linear OON structure, making the bonding
situation more stable in this case (see Figure S7).
Bending of the linear structures leads to a transformation of the
in-plane antibonding orbitals of the $\pi$-system (``a'' in Figure
\ref{fig:mo}A) into a non-bonding $p$-orbital on the oxygen at $\theta
= 90^\circ$.  The two non-bonding orbitals of the linear $\pi$-systems
transform also into $p-$orbitals on the oxygen.  Hence, at
$\theta=90^\circ$ three natural orbitals close in energy with mostly
$p-$orbital contribution on the oxygen atom arise. Their energy
fine-ordering depends on the amount of residual antibonding character
they bear. The out-of-plane antibonding orbital of the linear
$\pi-$systems (``b'' in Figure \ref{fig:mo}A) however transform into
an antibonding $\pi^*$ NO-orbital upon bending. Finally, the
antibonding orbital with dominant $\sigma^*$-character for the linear
structures ("c" in Figure \ref{fig:mo}A) also transforms into an
antibonding $\pi^*$ NO-orbital at $90^\circ$, considerably lowering
its orbital energy. The fine-ordering of the two $\pi^*$ NO-orbitals
again depends upon their remaining additional antibonding
character. Thus, for a T-shaped structure ($\theta=90^\circ$) the
quasi-degeneracies lead to a large number of configurations with
similar energy and lead to small energy differences for the eight
states included in the CASSCF wavefunction (see Figure
\ref{fig:mo}B). \\

\noindent
Two additional interesting observations on the NO+O channel can be
made from the MO diagram: 1) No stable covalent bonding between the
oxygen and the N-O fragment in the T-shaped structure is observed at the 
CASSCF level of theory. This explains the almost fully repulsive
character of the NO+O channel along $R$ for $\theta$ close to
90$^\circ$ (cf.\ Figure \ref{fig:1dpes}). 2) Upon bending, the
in-plane $\pi_3$* orbital (``a'' in Figure \ref{fig:mo}A)
significantly lowers its energy. As the
$\pi_3$* orbitals are partially occupied for the linear structures,
bending makes lower energy configurations accessible, yielding minima
on the PESs of the NO+O channel for slightly bent structures
(cf.\ Figure \ref{fig:pes}).  \\

\noindent
The CASSCF energies for the eight states along the bending coordinate
are shown in Figure \ref{fig:mo}B. In the following significant
features (i to ix) of the PESs are discussed. For the linear
structures (OON $(\theta =0)$ and ONO $(\theta = 180^\circ)$) the
orbital degeneracy leads to $^{2,4}$A$'$ and $^{2,4}$A$''$ lowest
states of equal energy (see points i to iv in Figure
\ref{fig:mo}B). Bending away from the linear geometry leads to an
approach and avoided crossing of the 1$^2$A$''$ and 2$^2$A$''$ states
($\theta=30^\circ$, point v), each of which is described by one
dominant configuration outside the crossing region. A similar
observation is made for the 1$^2$A$'$ and 2$^2$A$'$ states
($\theta=50^\circ$, point vi). The 1$^2$A$'$ state has a strong
multi-reference character with various configurations contributing in
an extended region $50^\circ \le \theta \le 100^\circ$. As indicated in
Figure \ref{fig:mo}A, the quasi degeneracies in the T-shaped
structures gives rise to a large number of configurations with similar
energies and to seven states within 0.9 eV for $\theta=90^\circ$
(point vii). The characteristics of points vii and ix can be described
along the same lines as for points vi and v, respectively.  The full
and detailed analysis of changes in configurations of the states along
the path is given in Figure S6.  It is noted that the
two lowest $^4$A states do not show an avoided crossing and have each
one dominant configuration along $\theta$. This explains the rather
simple topology of the 1$^4$A' PES in Figure \ref{fig:pes}.  The
inset in Figure \ref{fig:mo} amplifies the subtle changes in state
order around $\theta=90^{\circ}$. Various additional avoided crossings
can be observed (their analysis is given in Figure
S6.)\\

\subsection{Thermal Rates and Reaction Cross Sections}
\begin{figure}
\includegraphics[scale=0.3]{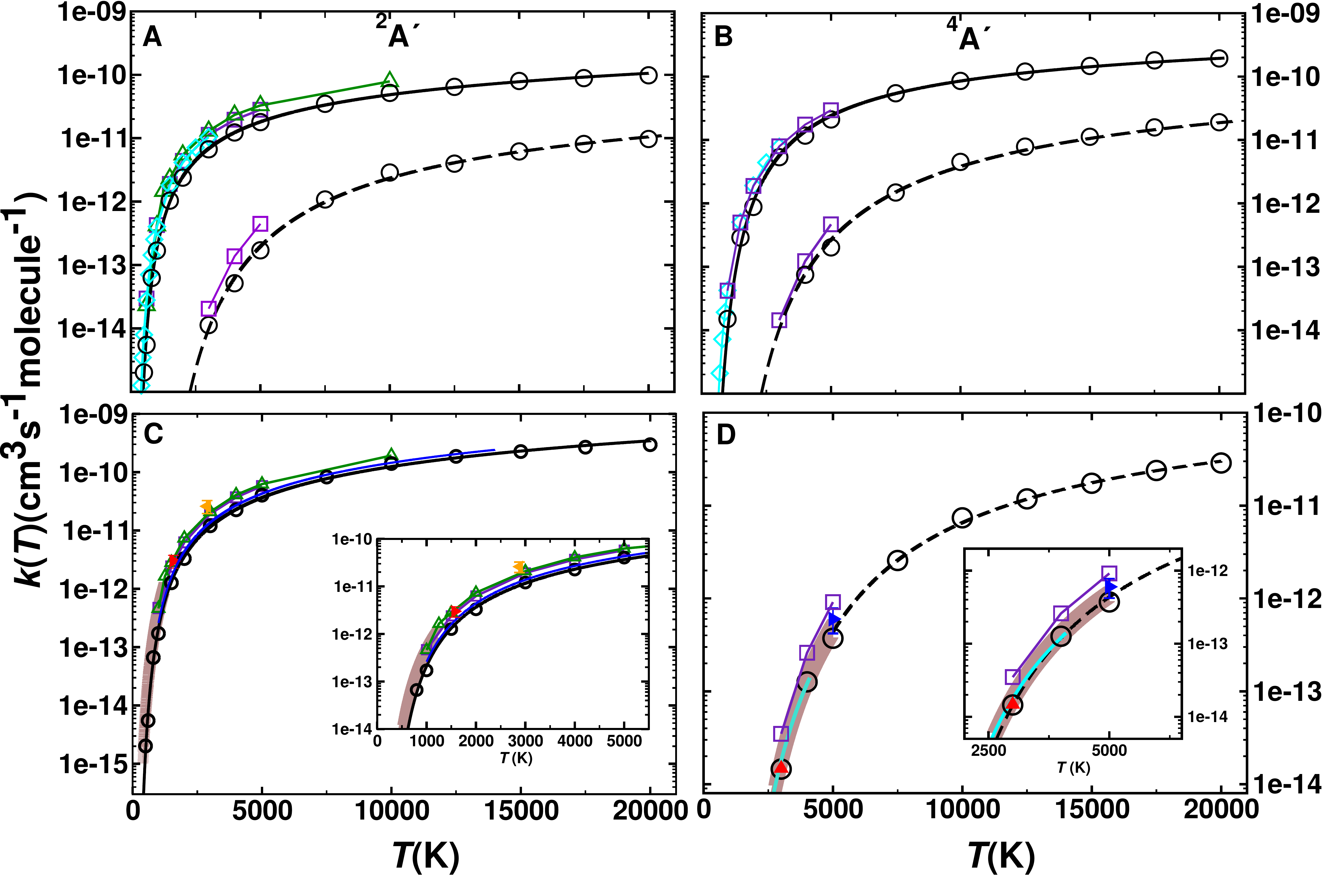}
\caption{(Top) Forward and reverse rate coefficients for the $^2$A$'$
  (panel A) and $^{4}$A$'$ (panel B) states. Rates for the forward
  (N($^4$S) + O$_2$(X$^3 \Sigma^-_g ) \rightarrow$ O($^3$P) +NO(X$^2
  \Pi$ ), open circles and solid black line) and reverse (O($^3$P) +
  NO(X$^2 \Pi) \rightarrow$ N($^4$S) +O$_2$(X$^3 \Sigma _g^-$) open
  circles and dashed lines) reaction are given separately. Results
  from previous computations based on VTST (green open
  triangle)\cite{Caridade2004}, ICVT (violet open square)\cite{Says2002}
  and quantum treatments (cyan open diamond)\cite{sultanov:2006} are
  also shown for comparison.  (Bottom) Total rates $k(T)$ for the
  forward (N($^4$S) + O$_2$(X$^3 \Sigma^-_g) \rightarrow$ O($^3$P)
  +NO(X$^2 \Pi$), panel C) and reverse (O($^3$P) + NO(X$^2 \Pi)
  \rightarrow$ N($^4$S) +O$_2$(X$^3 \Sigma _g^-$), panel D)
  reaction. The black open circles are the data from GB and the fit to
  a 3-parameter Arrhenius model is the solid black line. The fitting
  parameters are reported in Table \ref{tab:fit}. Results from VTST
  (green open triangle up)\cite{Caridade2004}, ICVT (violet open
  square)\cite{Says2002}, quantum (cyan open diamond)\cite{sultanov:2006}
  and evaluation (blue solid line)\cite{Bose1997}. Experimental values are also reported in Panel C ((red solid right triangle)\cite{Kaufman:1959}, (orange solid left triangle)\cite{Livesey:1971}) and Panel D ((blue solid right triangle)\cite{Wray:1962}, (red solid triangle up)\cite{Clark:1969}, (blue cyan line)\cite{Hanson:1974}) together with fits to experiment with errors (brown shaded areas in panels C\cite{Fernandez:1998} and D\cite{Hanson:1984}).}
\label{fig:rates}
\end{figure}

Thermal rates for the forward (N($^4$S) + O$_2$(X$^3 \Sigma^-_g )
\rightarrow$ O($^3$P) +NO(X$^2 \Pi$)) and reverse (O($^3$P) + NO(X$^2
\Pi) \rightarrow$ N($^4$S) +O$_2$(X$^3 \Sigma _g^-$)) reaction are
determined between 300 and 20000 K. A total of 50000 trajectories was
calculated at each temperature for each reaction on each electronic
states. The individual contributions of the $^2$A$'$ and $^4$A$'$
states are reported in Figure \ref{fig:rates} panels A and B. The
forward rates are about one order of magnitude higher than the reverse
rates for both electronic states. Comparison with previous
calculations\cite{Says2002} (see Figure S8) shows that
for both states and both directions they differ by a factor of $\sim
2$ for high temperature. For lower temperature they rather differ by a
factor of $\sim 5$. \\

\begin{tiny}
\begin{table}[h]
\begin{tabular}{l|cc|cc|ccc}
Parameter & $^{2}$A$'$ & Lit.\cite{Caridade2004} & $^{4}$A$'$ &
Lit.\cite{Caridade2004} & Total & Lit.\cite{Bose1997} & Lit.\cite{Says2002} \\ \hline
Forward & & & & \\ \hline $n$ & 0.83 & 0.63 & 0.56 & 0.97 & 1.18 & 1.18 & 1.60\\ 
$A$ & 3.58 & 34.0 & 117.2 & 2.31 & 0.370 & 0.414 & 0.014 \\ 
$B$[K] & 4105 & 4043 & 8722 & 7459 & 4090 & 4005 & 2894 \\ 
\hline 
\hline 
Reverse & & & & \\ 
\hline 
$n$ & 0.74 & -- & 0.64 & -- & 0.40 & -- & 1.51\\ 
$A$ & 1.77 & -- & 12.1 & -- & 190.3 & -- & 0.01 \\ 
$B$[K] & 19653 & -- & 23505 & -- & 24520 & -- & 19115\\ 
\hline
\end{tabular}
\caption{Arrhenius 3-parameter model (Eq. \ref{eq:arrhenius}) for $600
  \leq T \leq 20000$ K for the forward (N($^{4}$S) + O$_2$(X$^3
  \Sigma^-_g$) $\rightarrow$ O($^3$P) +NO(X$^{2} \Pi$)) and reverse
  (O($^{3}$P) + NO(X$^{2} \Pi$) $\rightarrow$ N($^4$S) +O$_{2}$(X$^3
  \Sigma_g^-$)) reaction. $A$ in units of $10^{-14}$ cm$^3$/(s
  molecule).}
\label{tab:fit}
\end{table}
\end{tiny}

\noindent
Figures \ref{fig:rates} C and D show the total rate $k(T)$ for the
forward and reverse reaction. The total rate is calculated by summing
the contributions from $^2$A$'$ and $^4$A$'$ surfaces. For practical applications, such
as discrete sampling Monte Carlo (DSMC) simulations,\cite{dsmc} it is
also useful to fit the data to an empirical, modified Arrhenius
relationship
\begin{equation}
    k(T)=A \cdot T^{n} \cdot e^{(-\frac{B}{T})}
\label{eq:arrhenius}
\end{equation}
The fitted parameters are given in Table \ref{tab:fit}. Additional
fits for N+O$_2$ on the $^4$A$'$ state yield\cite{Caridade2004}
$A=1.41$ $10^{-14}$ cm$^3$/(s molecule), $n=1.04$, $B=6112$ K based on
VTST data.\cite{Says2002} Fitting of earlier QCT data\cite{Bose1997}
yields remarkably similar values to the present results, see Table
\ref{tab:fit} and blue trace in Figure \ref{fig:rates}C, which,
however, differ both substantially from a more recent
study.\cite{Says2002} Experimental rates at higher temperatures are
rare. One study was carried out at 1575 K\cite{Kaufman:1959} which is
in quite good agreement within typical\cite{Hanson:1984} uncertainties
of 25 \% with the present simulations (Figure \ref{fig:rates}C)
whereas the rate from an experiment at higher temperature (2880
K)\cite{Livesey:1971} is larger than the rate from the present and
earlier\cite{Bose1997} simulations by about a factor of two. One
possible explanation is that for experiments above $T \sim 2000$ K
there is interference between the O+N$_2$ and N+O$_2$ reactions and
the analysis required a reaction network both of which introduce
uncertainties in the rate.\cite{Livesey:1971} For the reverse rate the
present simulations accurately describe those measured
experimentally.\cite{Wray:1962,Clark:1969,Hanson:1984}\\

\noindent
The reverse rate (O+NO) in Ref.\cite{Caridade2004} was not determined
from QCT simulations but rather by first computing the equilibrium
constant $K_{\rm eq}(T)$ according to statistical mechanics and then using
$k_{-}(T) = k_{+}(T)/K_{\rm eq}(T)$.  The Arrhenius values from
Ref.\cite{Caridade2004} are $A=0.114$ $10^{-14}$ cm$^3$/(s molecule),
$n=1.13$, and $B=19200$ K. To the best of our knowledge the present
work determined $k_{-}(T)$ for the first time from QCT simulations.\\
  
\begin{figure}[h]
\centering \includegraphics[scale=0.6]{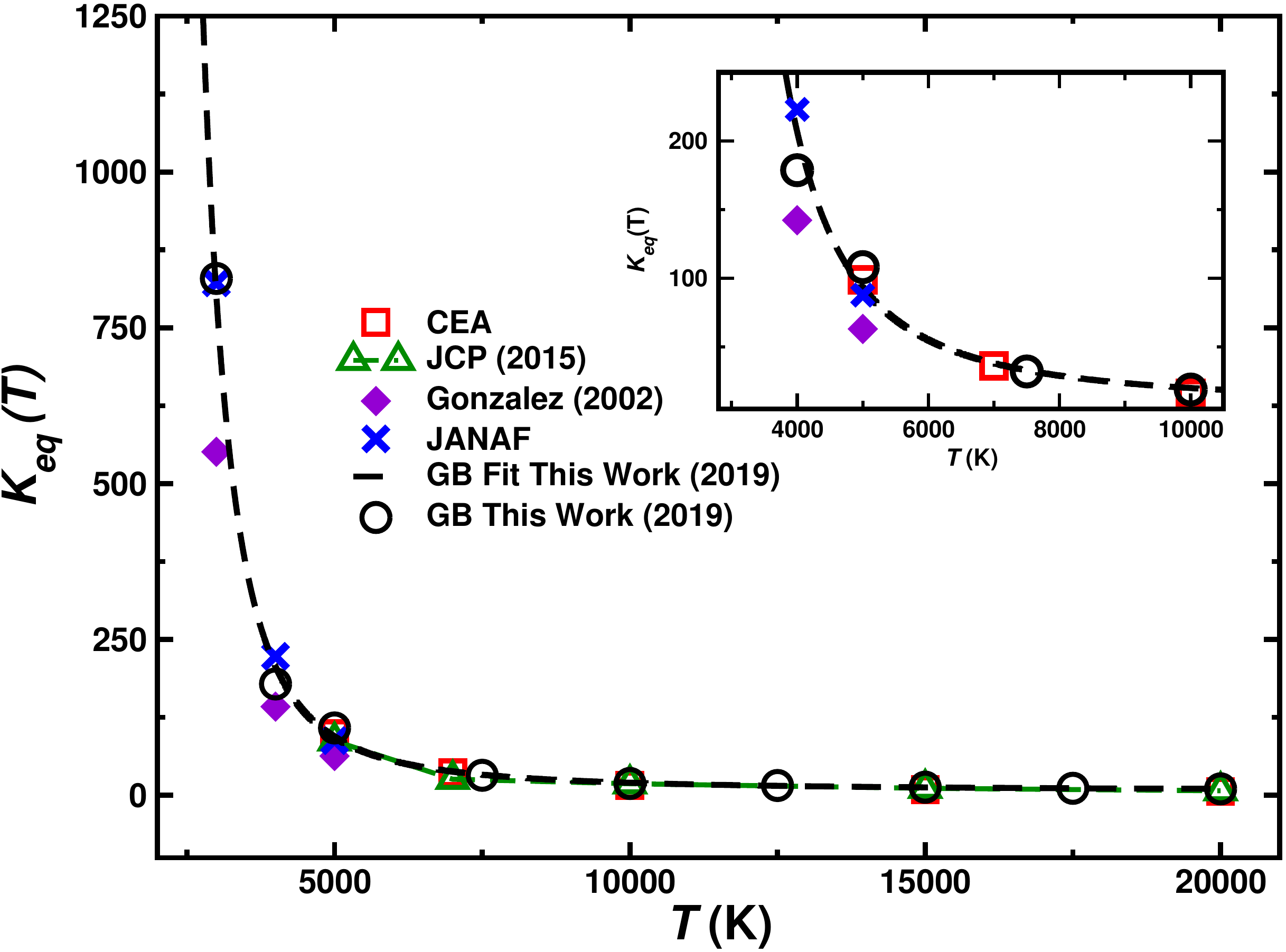}
\caption{Equilibrium constant $K_{eq}(T)$.  QCT results calculated in
  this work (circles), fit to a modified Arrhenius model (solid lines)
  for temperatures between 1000 and 20000 K.  Previous
  QCT\cite{CastroPalacio2015}, JANAF tables \cite{Chase1985} and
  Chemical Equilibrium with Application (CEA) results \cite{CEA1996}
  are also included for comparison. The inset shows an enlarged view
  for lower temperatures.}
\label{fig:econstant}
\end{figure}

\noindent
In addition, the equilibrium constant $K_{eq}(T)$ as defined in Eq. \ref{eq:equilibrium} is also calculated  (see Figure \ref{fig:econstant}) as it can be compared directly with experimental work. For $K_{\rm eq}(T)$ the present calculations agree favourably with the JANAF and CEA values
over the entire temperature range, as can be expected since $K_{\rm
  eq}$ is also determined from the difference in Gibbs free energy
between the initial and final states.\\

\noindent
To determine the cross sections depending on the final vibrational
state $v'$, additional simulations were carried out. For this $2
\times 10^5$ independent trajectories on each of the three electronic
states were run starting from the N+O$_2$ asymptote with a
distribution of O$_2$ internal $(v,j)$ states at 1000 K to follow
N($^4$S) + O$_2$(X$^3\Sigma^-_g$) $\rightarrow$ O($^3$P)+
NO(X$^2\Pi)(\nu^{'},j)$.  Then, total reaction cross sections for the
individual final vibrational states $(\nu',j)$ were determined. The
cross section as a function of the vibrational level ($\nu'$) is
reported in Table \ref{tab:vibcross} and compared with previous
experimental\cite{Caledonia:2000} and theoretical\cite{Caridade2004}
work. Of particular interest is the dependence of $\sigma$ on the
final vibrational state $\nu '$ of NO because experimentally, an
oscillating total cross section had been found with minima at $\nu' =
3$ and $\nu' = 5$.\cite{Caledonia:2000} However, earlier
experiments\cite{Rahbee:1981,Herm:1983} report the rate constant for
formation of NO for the N + O$_{2}$ $\rightarrow$
NO + O reaction for vibrational levels $\nu=2-7$. Using
Eq. \ref{eq8} these rates were converted into cross sections which are
monotonically decreasing with $\nu$ except for
$\nu=2$.\cite{Rahbee:1981} Rates for $\nu=0$ and $\nu=1$ were reported
to be larger compared to $\nu \geq 2$.\cite{Rahbee:1981} A comprehensive
comparison of the present results for the cross sections is given in
Table \ref{tab:vibcross}.\\

\begin{tiny}
\begin{table}[h!]
\begin{center}
\begin{tabular}{c|cc|ccc|ccc}
  State $\nu'$ & Exp.\cite{Caledonia2000} & Exp.\cite{Rahbee:1981} & $^2$A$'$ & $^4$A$'$ & Total & Lit.\cite{Duff1994} & Lit.\cite{Ramachandran2000} & Lit.\cite{Caridade2004}  \\
  \hline
0 & -- & -- & 0.17 & 0.67 & 0.84    & ---  & 0.28 & 0.50 \\
1 & 0.49 & -- & 0.16 & 0.48 & 0.64  & 0.37 & 0.41 & 0.53 \\
2 & 0.65 & 0.37 & 0.14 & 0.31 & 0.45  & 0.42 & 0.46 & 0.48 \\
3 & 0.20 & 0.39 & 0.14 & 0.22 & 0.36  & 0.39 & 0.44 & 0.43 \\
4 & 0.69 & 0.22 & 0.12 & 0.16 & 0.28  & 0.36 & 0.37 & 0.32 \\
5 & 0.37 & 0.16 & 0.10 & 0.10 & 0.20  & 0.31 & 0.32 & 0.26 \\
6 & 0.20 & 0.04 & 0.09 & 0.07 & 0.16  & 0.27 & 0.28 & 0.20 \\
7 & 0.25 & 0.03 & 0.07 & 0.05 & 0.12  & 0.23 & 0.24 & 0.18 \\ 
\end{tabular}
\caption{Individual and total cross sections (in \AA\/$^{2}$) for the
  N + O$_{2}(\nu)$ $\rightarrow$ NO($\nu'$) + O process as a
  function of the final vibrational state ($\nu'$) from Gaussian
  binning. The total cross sections are also compared with
  experimental results (Exp.)\cite{Caledonia2000} and rates for NO
  formation (Exp.)\cite{Rahbee:1981} converted to cross sections
  according to Eq. \ref{eq8}.  Additionally, comparison with other
  computational work
  (Lit.).\cite{Caridade2004,Duff1994,Ramachandran2000} is also
  provided.}
\label{tab:vibcross}
\end{center}
\end{table}
\end{tiny}
  
\begin{figure}[h]
\centering \includegraphics[scale=0.6]{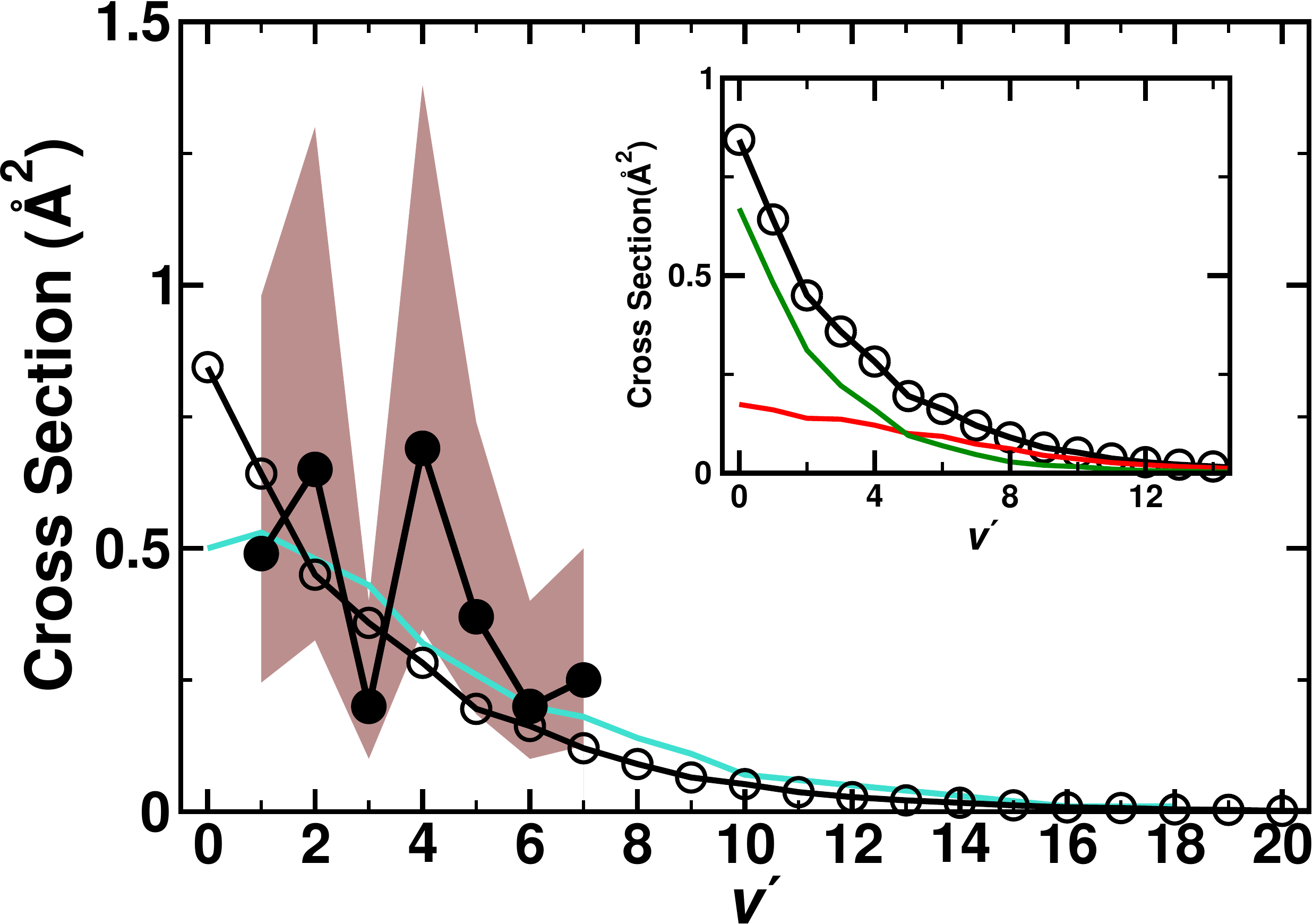}
\caption{Vibrational state-dependent, total cross sections (in
  \AA\/$^2$) for the N + O$_{2}(\nu)$ $\rightarrow$ NO($\nu'$)
  + O process as a function of the final $\nu'$. Total contribution
  ($^2$A$'$ $+$ $^4$A$'$) (open circles and black line) compared with
  previous computations (turquoise line)\cite{Caridade2004} and
  experiment (black solid circles)\cite{Caledonia:2000} with error
  region (brown shaded area) results. The inset reports the individual
  contributions from the present work for the $^2$A$'$ (red) and
  $^4$A$'$ (green), together with the total cross section.}
\label{fig:statecross}
\end{figure}

\noindent
No oscillating behaviour of the cross section was found from the
present simulations, see Figure \ref{fig:statecross}. This finding
agrees with previous simulations\cite{Caridade2004} based on a DIM PES
for the $^2$A$'$ state\cite{varandas:2003} and a fitted MBE to CASPT2
calculations for the $^4$A$'$ state.\cite{Says2002} The present
calculations find a decaying cross section with higher vibrational
state. The individual contributions of the ($^2$A$'$) and ($^{4}$A$'$)
state are calculated in addition to the total cross sections. The previous computational work\cite{Caridade2004}
also used the contribution of both the $^2$A$'$ and $^4$A$'$ states
and found a small population inversion peaking at $\nu' = 1$. But
overall, the findings from both simulation studies are consistent and
suggest that the experimental findings\cite{Caledonia:2000} should be
reconsidered.\\

\subsection{Vibrational relaxation}
As a third observable reactive (Oxygen exchange) and non-reactive
vibrational relaxation of O + NO($\nu=1$) $\rightarrow$ O +
NO($\nu=0$) was studied on the RKHS PESs for all three electronic
states as a function of temperature. An early
experiment\cite{Dodd1999} determined the rate for NO($\nu=1$)
vibrational relaxation by O atoms at room temperature. The
vibrationally excited NO and relaxer O atoms were formed using 355 nm
laser photolysis of a dilute mixture of NO$_2$ in an argon bath
gas. The reported total rate was $k_{\nu=1 \rightarrow 0}= 2.4 \pm
0.5$ $10^{-11}$ cm$^{3}$s$^{-1}$ at $T=298$ K. It was argued that this
value is 2 to 3 times lower than the generally accepted value of $K$
used in atmospheric modeling.\cite{Fernando:1979,Sharma:2001}
Subsequent QCT simulations\cite{Caridade2008} on the $^2$A$'$ and
$^2$A$''$ states, find a value of $k_{\nu=1 \rightarrow 0} (T=298 {\rm
  K})= 2.124 \pm 0.73 (10^{-11})$ cm$^3$s$^{-1}$. It should be noted
that the PESs for these two states are based on different
approaches. For the $^2$A$'$ PES it is based on a DIM
ansatz\cite{varandas:2003} whereas the $^2$A$''$ PES is a MBE fit to
CASPT2 calculations.\cite{Gonzlez2001} Even earlier calculations using
the $^2$A$'$ PES obtained a somewhat smaller rate of $k_{\nu=1
  \rightarrow 0} (T = 298 {\rm K})= 1.7 (10^{-11})$
cm$^{3}$s$^{-1}$.\cite{quack:1975}\\

\noindent
A more recent experiment\cite{Hwang2003} used a continuous wave
microwave source to generate oxygen atoms, combined with photolysis of
trace amounts of added NO$_2$ to produce vibrationally excited NO. The
rate for vibrational relaxation is $k_{\nu = 1 \rightarrow 0} (T=295
{\rm K}) = 4.2 \pm 0.7 (10^{-11})$ cm$^3$s$^{-1}$ which is an increase
by 75 \% compared with the earlier results.\cite{Dodd1999} Later QCT
simulations\cite{Caridade2018} based on the DIM PES for the $^2$A$'$
state\cite{varandas:2003} and a fitted DMBE PES based on 1681
MRCI/AVQZ calculations for the $^2$A$''$ state\cite{Mota2012} report a
value of $k_{\nu=1 \rightarrow 0}(T=298 {\rm K}) = 4.34 \pm 0.7
(10^{-11})$ cm$^3$s$^{-1}$. Another computational
study\cite{Schinke:2007} reported a value of $k_{\nu = 1 \rightarrow
  0} (T = 300 {\rm K})\sim 5 (10^{-11})$ cm$^{3}$s$^{-1}$.\\

\begin{figure}[h]
\centering
\includegraphics[scale=0.6]{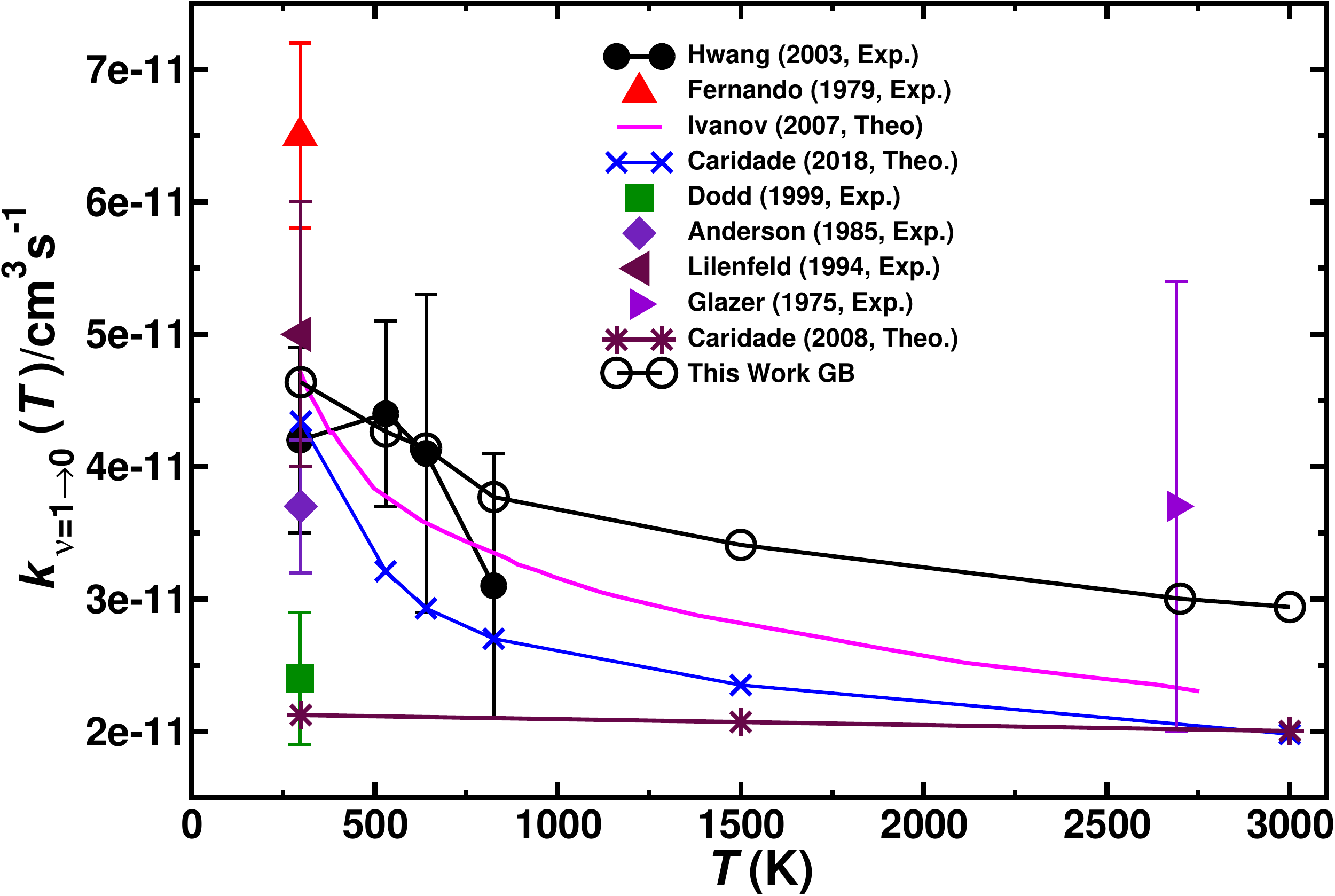}
\caption{Total vibrational relaxation rate for O+NO($\nu=1$)
  $\rightarrow$ O+NO($\nu=0$). The individual contributions of the
  $^2$A$'$, $^4$A$'$, and $^2$A$''$ states are provided in Figure
  S9. Present data from Gaussian binning are open black
  circles and literature values are the symbols as
  indicated.\cite{Hwang2003,Caridade2018,Dodd1999,Caridade2008,Ivanov2007,glanzer:1975,Fernando:1979,Lilenfeld:1994,Anderson:1985}}
\label{fig:vibrelax}
\end{figure}

\noindent
As to compare with the more recent experiments\cite{Hwang2003} the
individual contributions of the $^{2}$A$'$, $^{4}$A$'$, and
$^{2}$A$''$ states towards vibrational relaxation of O + NO($\nu = 1$)
$\rightarrow$ O + NO($\nu = 0$) were determined here. Additionally,
the total rate is compared with previous
theoretical\cite{Caridade2008,Caridade2018} and
experimental\cite{Hwang2003,Dodd1999} results, see Figure
\ref{fig:vibrelax}. In particular for low temperature the agreement
with the more recent\cite{Hwang2003} experiments and the only
high-temperature experiment (at 2700 K)\cite{glanzer:1975} is
noteworthy. The results from the simulations based on high-level,
2-dimensional PESs\cite{Ivanov2007} for the $^2$A$'$ and $^2$A$''$
states are also in good agreement with the experiments and the present
simulations.\\

\begin{tiny}
\begin{table}[h]
\begin{tabular}{c|lllllll}
 & 298 K & 530 K & 640 K & 825 K & 1500 K & 2700 K & 3000 K \\
\hline
$^2$A$''$  & 2.21  & 1.98  & 1.82   & 1.63  & 1.42 & 1.13 & 1.16 \\
$^4$A$'$   & 0.00  & 0.00  & 0.00  & 0.00  & 0.01 & 0.08 & 0.12 \\
$^2$A$'$   & 2.43  & 2.28  & 2.32  & 2.15  & 1.98 & 1.80 & 1.66 \\
\hline
Total           & 4.64  & 4.26  & 4.14  & 3.77  & 3.41 & 3.00 & 2.94 \\
\hline
Exp.\cite{Hwang2003} & $4.2 \pm 0.7$ & $4.4 \pm 0.7$ & $4.1 \pm 1.2$ & $3.1 \pm 1.0$ & -- & -- & -- \\
\end{tabular}
\caption{Electronic state-dependent vibrational relaxation rates (in
  units of $10^{11}) k_{\nu \rightarrow \nu'}$: O+NO($\nu=1$)
  $\rightarrow$ O+NO($\nu'=0$) for the $^{2}$A$'$, $^{4}$A$'$ and
  $^{2}$A$''$ states and the total contribution using GB.}
\label{tab:vr}
\end{table}
\end{tiny}

\noindent
As can be seen in Figure S9 and Table \ref{tab:vr},
trajectories run on the $^2$A$'$ state contribute most to the
vibrationally relaxing (VR) rates. Hence, to explore whether and how
the process of vibrational relaxation (O+NO($\nu=1$) $\rightarrow$
O+NO($\nu'=0$)) and sampling of the underlying PES are related,
another 25000 independent trajectories were run for the $^2$A$'$ state
at 530 K. Out of those, 5722 relaxed to $\nu'=0$ whereas for 13311
trajectories either the internal state of NO was changed to ($\nu'\neq
0,j'$) or O$_2$ was produced. For the remaining 5967 (nonreactive)
trajectories the initial ro-vibrational state is not changed. All the
trajectories were saved and rigorous analysis have been carried out to
investigate the relaxation process. The opacity function $P(b)$ for relaxing
trajectories computed on $^2$A$'$ PES is reported in Figure S10.\\

\begin{figure}[h!]
\centering
\includegraphics[scale=0.8]{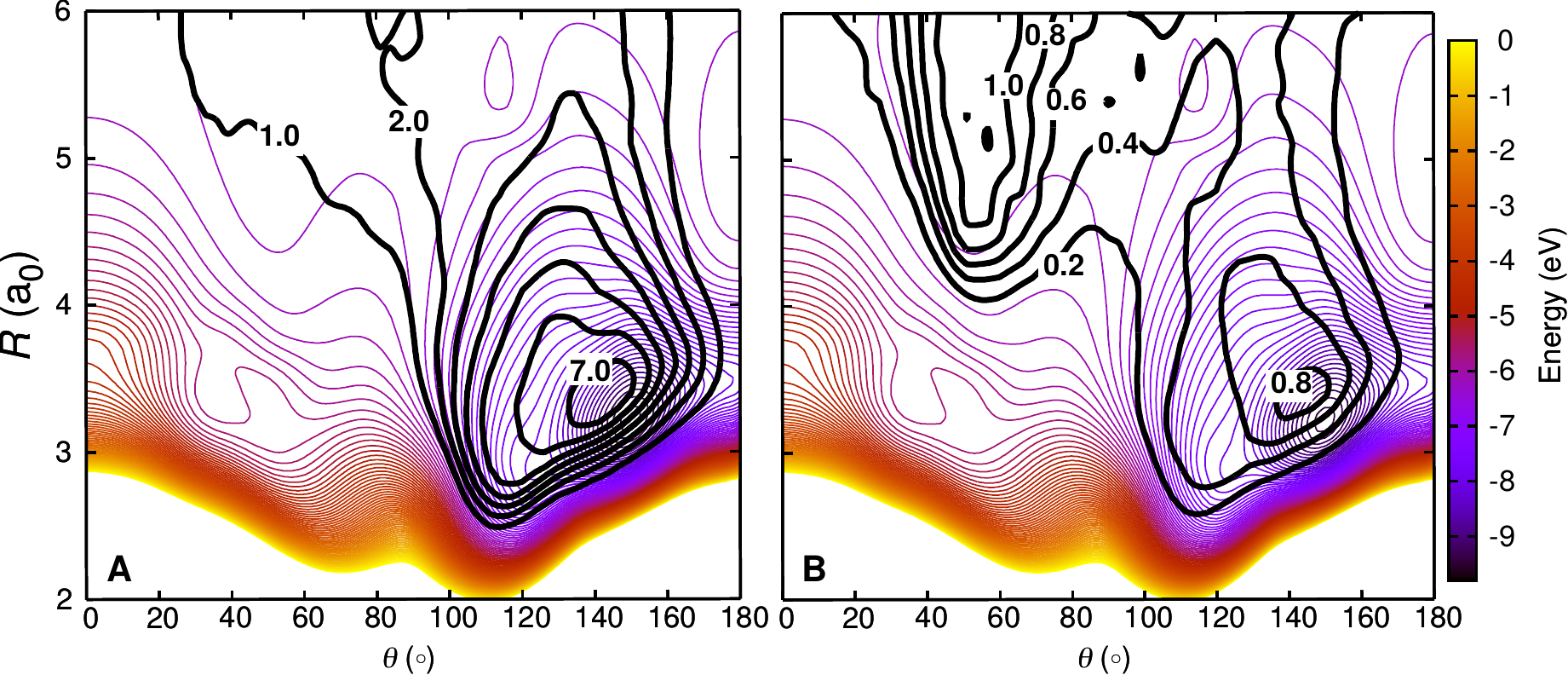}
\caption{Density trajectory map for the vibrationally relaxing (left)
  and non-relaxing (right) for O+NO($\nu=1,j$) collisions on the
  $^2$A$'$ PES. Vibrationally relaxing trajectories includes both,
  reactive and non-reactive trajectories, i.e. O$_{\rm A}$+NO$_{\rm
    B}$ $\rightarrow$ O$_{\rm B}$+NO$_{\rm A}$ and O$_{\rm
    A}$+NO$_{\rm B}$ $\rightarrow$ O$_{\rm A}$+NO$_{\rm B}$, whereas
  for the non-relaxing trajectories we excluded the trajectories for
  which the initial ro-vibrational state is not changed. The density
  map for the trajectories is superimposed on a relaxed 2D RKHS PES
  (see text for details). The two different classes of trajectories
  access different regions in configuration space, corresponding to
  different angular anisotropies.}
\label{fig:density}
\end{figure}

\noindent
The probability distributions of different O+NO configurations for
different types of trajectories have been computed in $(R, \theta)$
space. Structures are included in the computation only if any of the
NO bond is within 2.03 to 2.39 a$_0$ (the turning points for the
$\nu=1$ state of NO are $r_{\rm min}=2.046$ a$_0$ and $r_{\rm
  max}=2.370$ a$_0$). Gaussian binning with bin size $\Delta R = 0.1$
a$_0$ and $\Delta \theta = 3^\circ$ was used and contributions from
2.03 a$_0 < r_{\rm NO} < 2.39$ a$_0$ are excluded. Individually
normalized distributions for relaxing (Figure \ref{fig:density}A) and
nonrelaxing (Figure \ref{fig:density}B) trajectories are then
projected on an $r-$relaxed 2D PES. This PES was computed by
determining the minimum energy for given $(R, \theta)$ with $r \in
[2.03,2.39]$. Such an $r-$relaxed PES is a more realistic way for this
comparison as it also incorporates the varying NO bond length during
the dynamics instead of restricting it to one specified value.\\

\noindent
Figure \ref{fig:density} demonstrates that the two families of
trajectories sample distinct regions of the interaction potential. The
VR trajectories have a high density in the deep potential well area
(dark blue) of the PES and sample mostly $\theta>90^\circ$
region. This suggests formation of a long lived, tightly bound
collision complex. However, the non-relaxing (NR) trajectories spend
less time in the potential well region and the density map is rather
flat, more uniformly distributed along the angular coordinate with
slightly larger sampling in the low-$\theta$ region.\\

\noindent
To check the initial (before collision) angular dependence of the
trajectories and role of long-range anisotropic interactions between
the atomic collider and the diatomic target, similar density maps like
Figure \ref{fig:density} have been computed for the VR and NR
trajectories only up to the time satisfying the criterion that the sum
of the three inter-nuclear distances is less than 9.5 a$_0$. Those are
shown for the NR trajectories in Figure S11 and for the
relaxing in Figure S12. It can be seen that at a
separation of $\sim 8.5$ a$_0$ the distribution $P(\theta)$ already
has ``structure'' for the NR trajectories and in that a large fraction
samples the range $\theta \sim 50^\circ$ while the NR trajectories
scarcely sample the high-$\theta$ region. However, for the relaxing
trajectories the distribution is much more even and lacks a specific
high-probability characteristic for a particular angle. Since the
low-$\theta$ region of the PES is repulsive, most of the trajectories
are reflected with only changing the rotational state of the NO and
resulting NR events.  A large fraction of those NR trajectories could
not even visit the short-range interaction region ($R<6.0$ a$_0$) and
they fly by from the target contributing twice (incoming and outgoing
trajectories) more in the density map which is obvious in Figure
S11.\\

\noindent
In Figure S13, ten randomly selected VR (red) and NR
(black) trajectories from each of the data set plotted in Figure
S11 and S12 are projected on similar 2D PES
as in Figure \ref{fig:density}. The dashed lines represent the
reactive (oxygen exchange or O$_2$ formation) trajectories.  It can be
seen that all VR trajectories sample the potential well region which
supports a collision complex. Out of the 10 VR trajectories 3 involve
a reactive, oxygen exchange event. The ratio 7:3 is representative of
all trajectories (3890:1832, for relaxing non reactive vs. relaxing
reactive trajectories). Thus, oxygen exchange events contribute almost
one third to VR. On the other-hand, among the NR trajectories a
certain fraction accesses the global minimum of the PES but most of
them do not continue beyond $R<6.0$ a$_0$ but are reflected at longer
$R$.\\

\noindent
The results above suggest that relaxing and non-relaxing trajectories
probe different parts of the PES. Hence, in order to be able to
realistically describe vibrational relaxation the relevant regions,
especially the potential well of the PES, have to be described
sufficiently accurately. Figure S14 reports the same PES
together with the positions in $(R,\theta)$ for which MRCI+Q
calculations were carried out. It can be seen that the relevant
regions sampled by vibrationally relaxing and non-relaxing
trajectories are covered by the electronic structure
calculations. Thus the current PES is expected to provide an accurate
description of the interaction potential for relaxation dynamics,
which is also supported when comparing the computed rates with
experiments.\\

\section{Discussion and Conclusions}
QCT calculations were carried out on the $^{2}$A$'$, $^{4}$A$'$ and
$^{2}$A$''$ electronic states of NO$_2$ for both, the forward and
reverse reaction. The total rates agree favourably with experiment for
the forward and reverse reaction (Figures \ref{fig:rates}C and D), except for the experiment for the forward rate at 2880 K for which interference with other reactions render the analysis more difficult.\cite{Livesey:1971} The $T-$dependent
equilibrium constants are close to those reported in the JANAF tables
\cite{Chase1985} and to those from results reported in Chemical
Equilibrium with Application (CEA)\cite{CEA1996}. This latter fact
suggests that the forward rate $k_{+}(T)$ is in fact preferred over
the single available experimental result at higher temperature (2880
K).\cite{Livesey:1971} Vibrational relaxation rates were computed for
the O + NO($\nu=1$) $\rightarrow$ O + NO($\nu=0$) process.  Both
states, $^{2}$A$'$ and $^{2}$A$''$, contribute to vibrational
relaxation whereas the contribution from the $^{4}$A$'$ state is small
at low temperature ($k \approx 10^{-14}$) but increases for higher
temperatures (Table \ref{tab:vr}).\\

\noindent
For VR to occur, the force on the NO oscillator must act along the
chemical bond, not orthogonal to it. Hence, the PES along the
$\theta=0$ and $\theta=180^\circ$ directions are most relevant to
convert translational energy of the oxygen atom into relaxation of the
vibrational motion of the NO diatomic, see Figure \ref{fig:density}. As around $\theta=0$ the PES is
repulsive it is primarily the region around $\theta=180^\circ$ to
which VR is sensitive to. The present work highlights that different parts of the PESs are
probed depending on the observable considered, which can even be demonstrated
explicitly. For example, using the DIM PES for the $^2$A$'$ state for
computing the N+O$_{2}\rightarrow$O+NO temperature-dependent rate
coefficients together with contributions for the $^4$A$'$ state from
the literature, acceptable agreement with experiment can obtained
whereas for the temperature dependent vibrational relaxation the DIM
PES finds a $T$-independent rate (see Figure \ref{fig:vibrelax}) which considerably underestimates that reported from experiments.\\

\noindent
The fact that different observables provide information about
different parts of the PES has already been highlighted for van der
Waals complexes. As an example, the morphed PESs for the Ne--HF
complex\cite{mm.nehf:1998} demonstrated that observables from high
resolution spectroscopy about the lowest stretching and bending states
along the van der Waals coordinate provide sensitive information about
the linear Ne--HF approach but no information about the antilinear
Ne--FH part of the PES. Hence, it will be interesting to relate the
space sampled by trajectories leading to particular final states with
specific features such as to better understand what parts of a PES are
crucial for reliably characterizing experimental observables from
high-level computational studies.\\

\noindent
It is expected that the temperature dependence of the rates computed
in the present work extrapolate more reliably to higher temperature
than the experimental data because, as the collision energy increases,
the simulations sample the near vertical repulsive wall of the
diatomic, determining its size. As this is an exponentially increasing
curve, errors in the exponent will make little difference in the
radius that is accessible at a given energy.\\

\noindent
The present work uses one of the highest affordable levels of theory
for the electronic structure calculations (MRCI+Q) and the validity of their
representation as a RKHS is thoroughly tested using a large number of
off-grid points. No relevant crossings between the PESs were found
which would require the inclusion of nonadiabatic effects into the
dynamics as had been done for the [CNO] system.\cite{Koner2018} As
with previous work for which quantum and classical nuclear dynamics
studies were carried out and found to agree with one
another\cite{Koner2018}, no quantum effects are expected for the
present system.\\

\noindent
In summary, the reactive dynamics, thermal rates and vibrational
relaxation for the ${\rm N}(^4S) +{\rm O}_2(X^3\Sigma^-_g)
\leftrightarrow {\rm O}(^3P) + {\rm NO}(X^2\Pi)$ reaction on the three
lowest potential energy surfaces was studied based on QCT
simulations. The results are consistent with most of the available
experiments. This provides a solid basis for a molecularly refined
picture of vibrational relaxation and extrapolation of thermal rates
to higher temperatures relevant at the hypersonic flight regime which
can be used for more coarse grained studies such as DSMC
simulations.\\

\section{Acknowledgment}
Part of this work was supported by the United State Department of the
Air Force which is gratefully acknowledged (to MM). Support by the
Swiss National Science Foundation through grants 200021-117810, the
NCCR MUST (to MM), sciCORE cluster and the University of Basel is also
acknowledged.

\newpage
\bibliography{refs}

\end{document}


\title{Supporting information: \\ The ${\rm N}(^4S) +{\rm
    O}_2(X^3\Sigma^-_g) \leftrightarrow {\rm O}(^3P) + {\rm
    NO}(X^2\Pi)$ Reaction: Thermal and Vibrational Relaxation Rates
  for the $^{2}$A$'$, $^{4}$A$'$ and $^{2}$A$''$ States}

\author{Juan Carlos San Vicente Veliz}
\affiliation{Department of Chemistry, University of Basel,
Klingelbergstrasse 80, CH-4056 Basel, Switzerland}

\author{Debasish Koner} 
\affiliation{Department of Chemistry, University of Basel,
Klingelbergstrasse 80, CH-4056 Basel, Switzerland}

\author{Max Schwilk} 
\affiliation{Department of Chemistry, University of Basel,
Klingelbergstrasse 80, CH-4056 Basel, Switzerland}

\author{Raymond J. Bemish} \affiliation{Air Force Research Laboratory,
  Space Vehicles Directorate, Kirtland AFB, New Mexico 87117, USA}

\author{Markus Meuwly}
\affiliation{Department of Chemistry, University of Basel,
Klingelbergstrasse 80, CH-4056 Basel, Switzerland}
\email{m.meuwly@unibas.ch}

\maketitle

\begin{figure}[h]
\centering \includegraphics[scale=0.7]{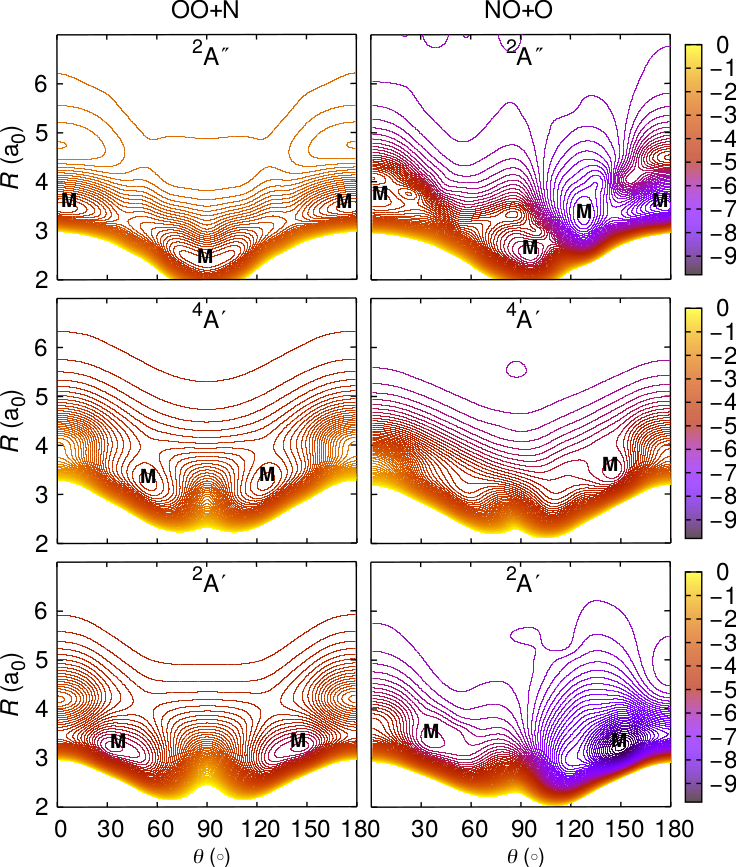}
\caption{Two-dimensional cuts through the 3-d PES for the OO+N (left
  and the NO+O (right) channels. The OO and NO diatomics are at their
  equilibrium bond lengths of the respective states, see text. The
  three states are $^{2}$A$'$, $^{4}$A$'$, and $^{2}$A$''$ from bottom
  to top. Contour increments of 0.1 eV. The zero of energy is for
  dissociation into atomic fragments. ``M'' labels the minima
  described in the text.}
\label{fig:sifig1t}
\end{figure}

\begin{figure}
\centering
\includegraphics[scale=0.7]{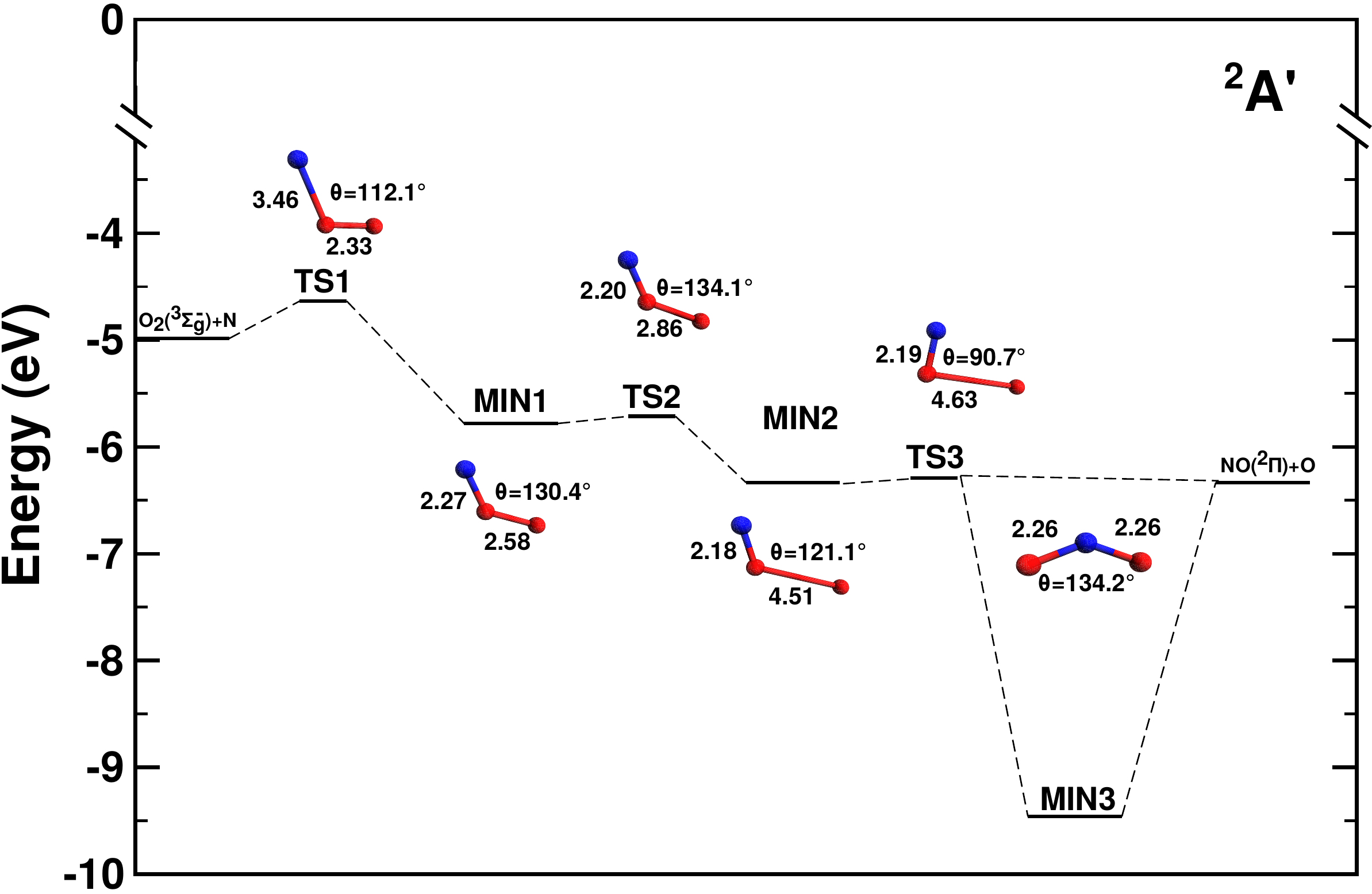}
\caption{The minima (MIN$i$) and transition states (TS$i$) for the
  $^2$A$'$ state as found from minimization and the nudged elastic
  band calculations.\cite{jonsson:2000,hammer:2016} The geometrical
  parameters are also given (bond distances in a$_0$).}
\label{fig:sifig2}
\end{figure}

\begin{figure}
\centering
\includegraphics[scale=0.7]{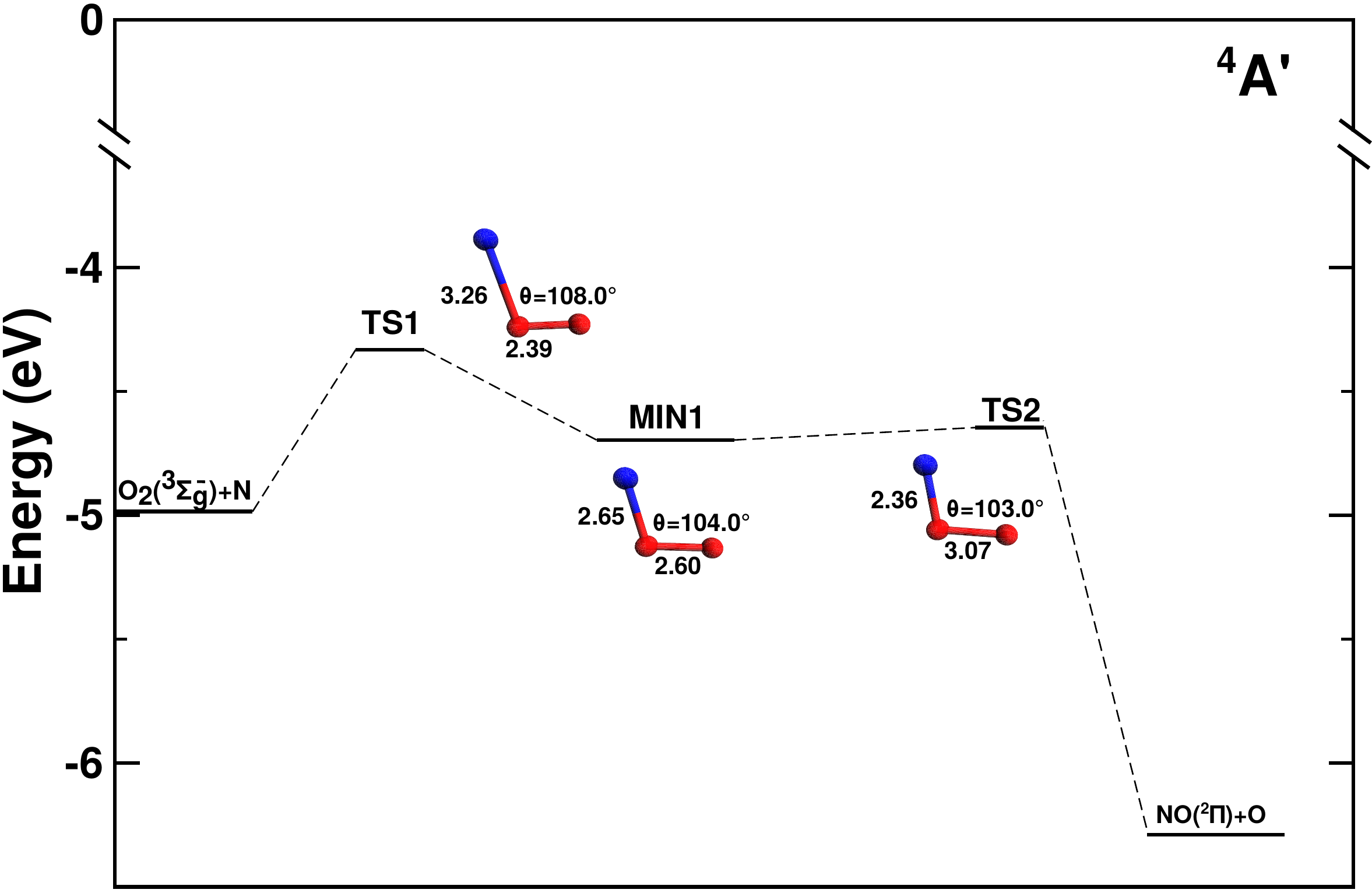}
\caption{The minima (MIN$i$) and transition states (TS$i$) for the
  $^4$A$'$ state as found from minimization and the nudged elastic
  band calculations.\cite{} The geometrical parameters are also given
  (bond distances in a$_0$).}
\label{fig:sifig3}
\end{figure}

\begin{figure}
\centering
\includegraphics[scale=0.7]{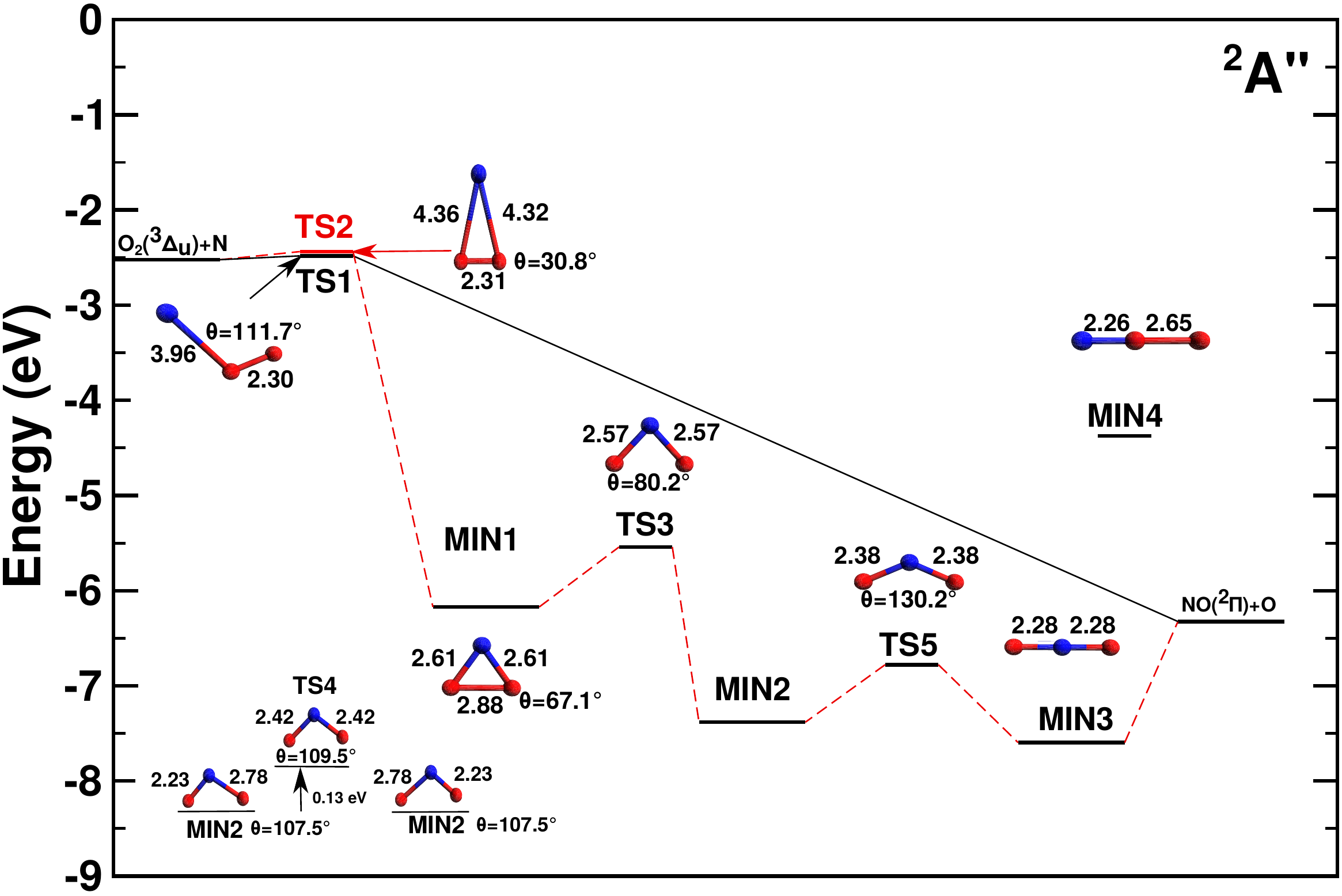}
\caption{The minima (MIN$i$) and transition states (TS$i$) for the
  $^2$A$''$ state as found from minimization and the nudged elastic
  band calculations.\cite{jonsson:2000} The geometrical parameters are also given
  (bond distances in a$_0$).}
\label{fig:sifig4}
\end{figure}

\begin{figure}
\centering
\includegraphics[scale=0.7]{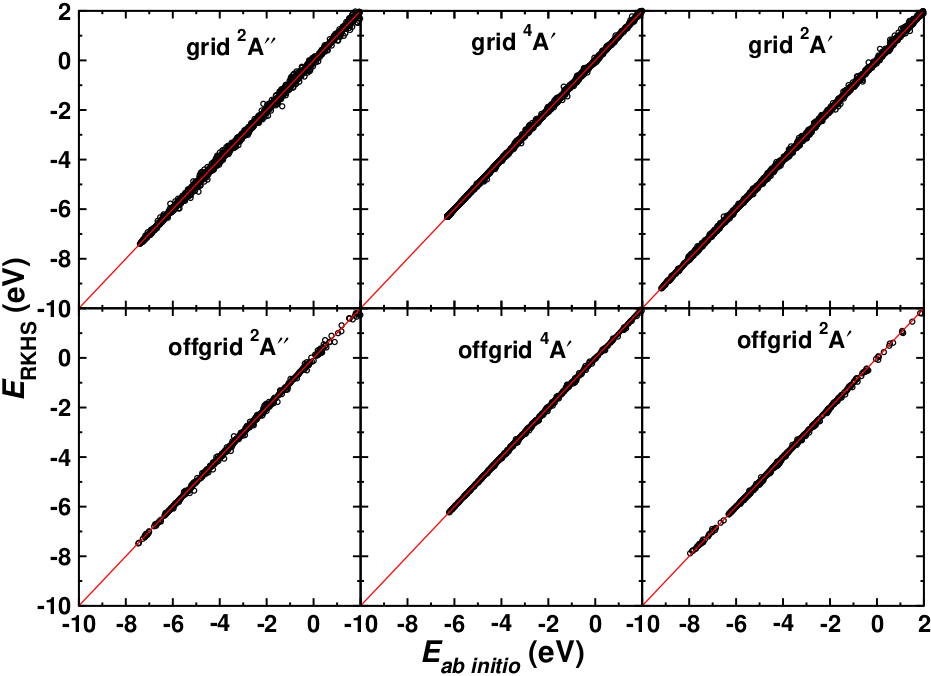}
\caption{Correlation between MRCI/aug-cc-PVTZ ($E_{\it ab initio}$)
  and RKHS energies up to a values of 2 eV for 7435 ($^2$A$'$), 6869
  ($^4$A$'$) and 7275 ($^2$A$''$) grid points and 537, 533 and 596
  offgrid points for the $^2$A$'$, $^4$A$'$ and $^2$A$''$ surfaces,
  respectively. The zero of energy is the O+O+N dissociation
  limit. The $R^2$ value for the grid points are (0.99984, 0.99989,
  0.99965) and for off-grid points (0.99959, 0.99966, 0.99922) for the
  ($^2$A$'$, $^4$A$'$, $^2$A$''$) surfaces, respectively. The
  corresponding root mean squared errors (RMSE) for the $^2$A$'$,
  $^4$A$'$ and $^2$A$''$ surfaces are (0.65, 0.49, 0.99) kcal/mol
  (0.028, 0.022, 0.043) eV for the grid points and (0.86, 0.76, 1.31)
  kcal/mol (0.038, 0.033, 0.057) eV for offgrid points.}
\label{fig:sifig1}
\end{figure}

\begin{figure}[h!]
\includegraphics[scale=0.04]{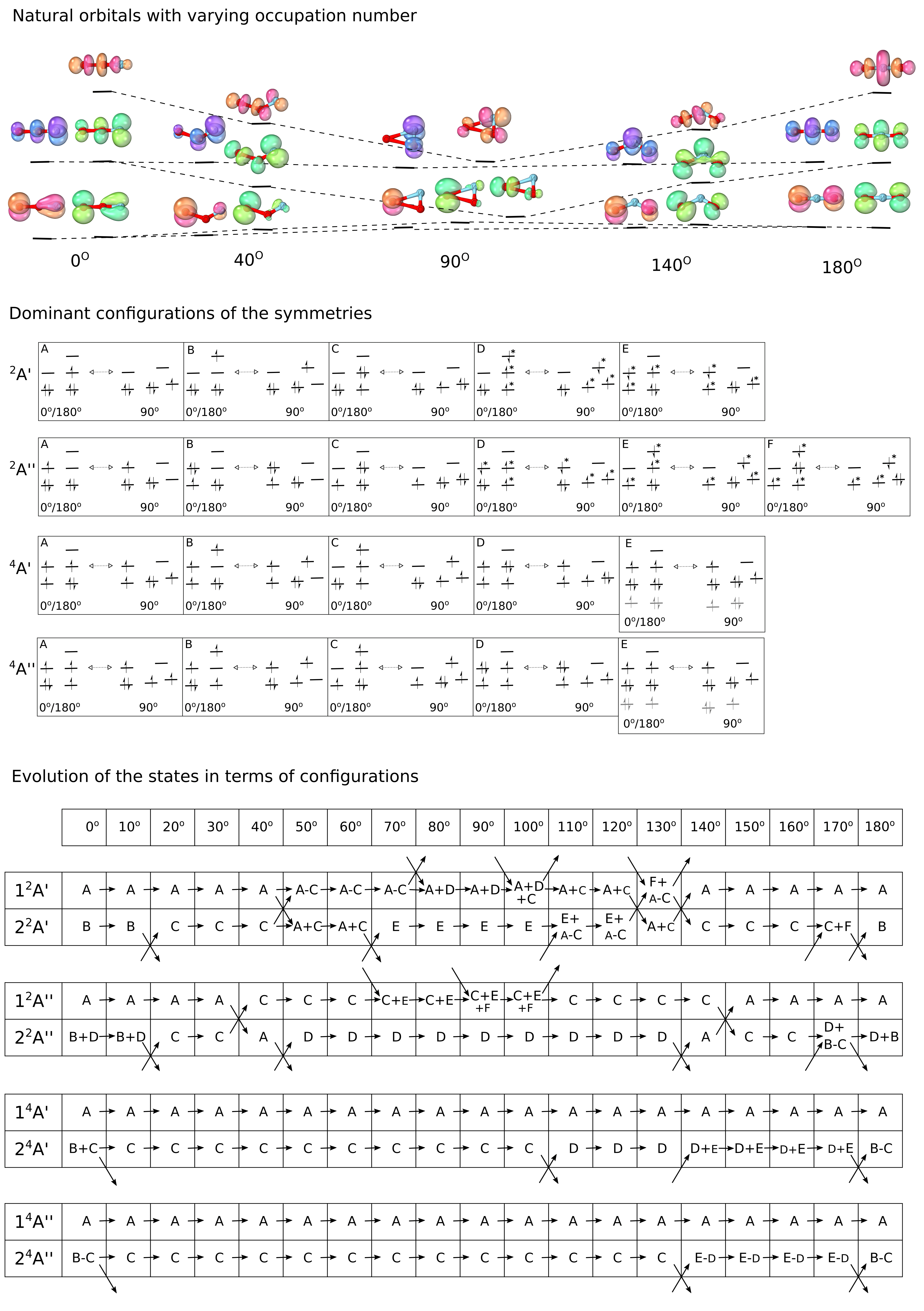}
\caption{MO diagram of the NO +O channel at fixed values of $R$ and
  $r$ together with details of the orbital occupancies.  This Figure
  complements Figure 3 in the main text. The most important
  configurations with $>$10\% contribution to the CASSCF wave function
  for each state are given in the middle panel of the figure.
  \emph{Caption continued on next page.} }
\label{fig:sifig6t}
\end{figure}

\addtocounter{figure}{-1}
\begin{figure} [h!]
  \caption{\emph{Continued:} Configurations where three single
    occupancies in orbitals couple to an overall doublet state are not
    distinguished further and so coupled orbitals indicated by an
    asterisk. Two configurations involve partially unoccupied bonding
    orbitals of the $\pi_3$-systems (orbitals shown in Figure
    \ref{fig:sifig7t}). In these extended MO diagrams, the additional
    orbitals are given in gray colour.  The bottom panel reports which
    of the configurations contribute to the states for a given bending
    angle $\theta$ according to the CI coefficients from the CASSCF
    wave function (labels A to F).  Smaller font indicates less
    contribution from a certain configuration. Arrows indicate the
    evolution of a configuration contribution to a state. I.\ e.,
    crossing arrows indicate a change of configuration of a state,
    normally resulting in a configuration coupling and an avoided
    crossing of two states. States $^2$A$'$ and $^2$A$''$ are
    characterized by strong changes in contributing configurations
    along the path which is at the origin of the complexity of the
    shape of the PESs. 
    Contrary to that, the
    $^4$A$'$ state only contains one configuration along the path which explains its
    rather simple topology.}
\end{figure}

\begin{figure}[h!]
\centering
\includegraphics[scale=0.05]{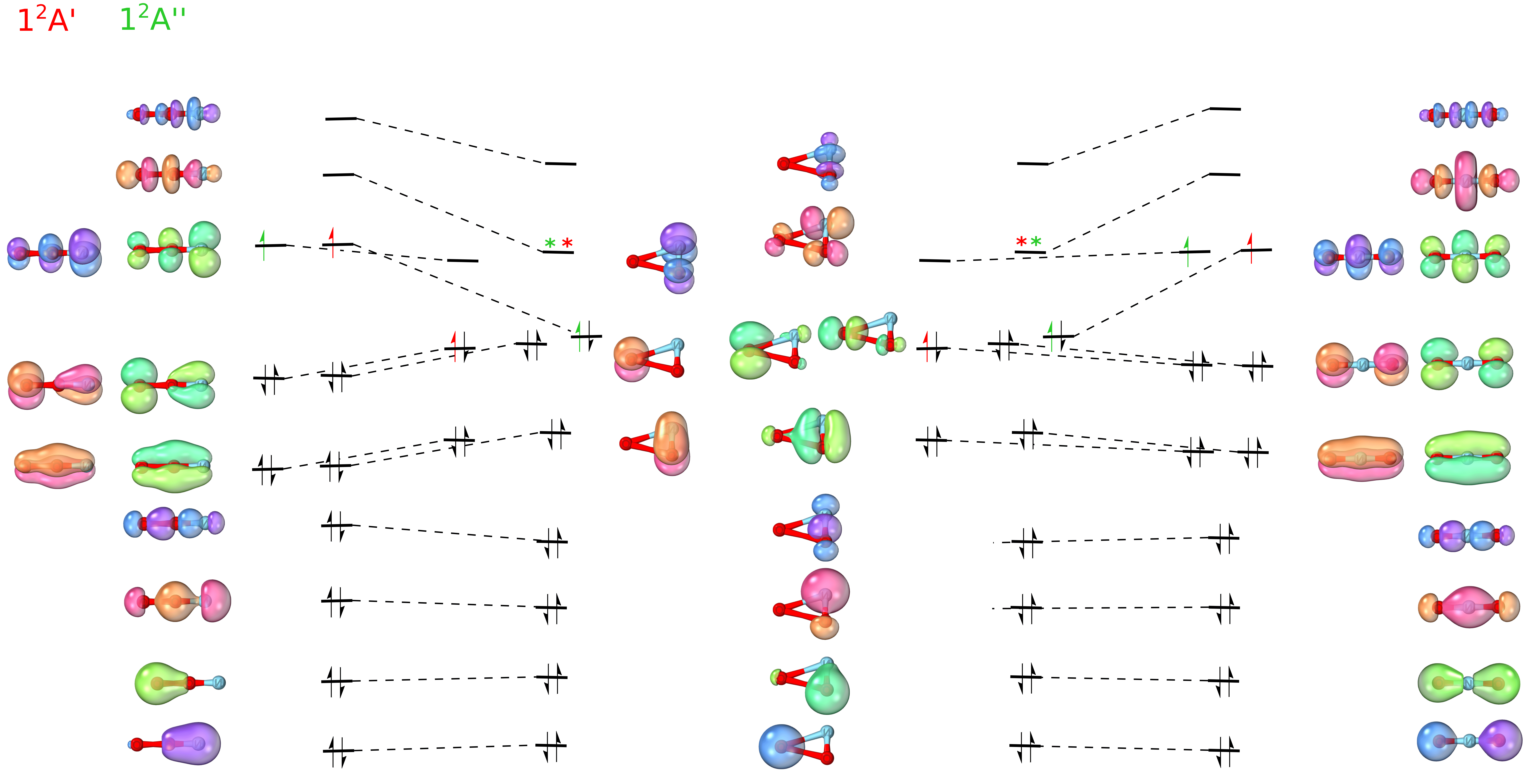}
\caption{MO diagram of NO$_2$ for the doublet ground state for the two
  linear configurations (left OON, right ONO) and for $\theta =90$
  (middle). The full valence orbital basis is shown. The dominant
  configurations at selected angles are depicted for the lowest
  $^2$A$'$ and $^2$A$''$ states. Black arrows for occupancies in both
  states, red for $^2$A$'$ and green for $^2$A$''$. Asterisks for
  significant additional occupancies due to strong correlation.}
\label{fig:sifig7t}
\end{figure}

\begin{figure}
\centering
\includegraphics[scale=0.30]{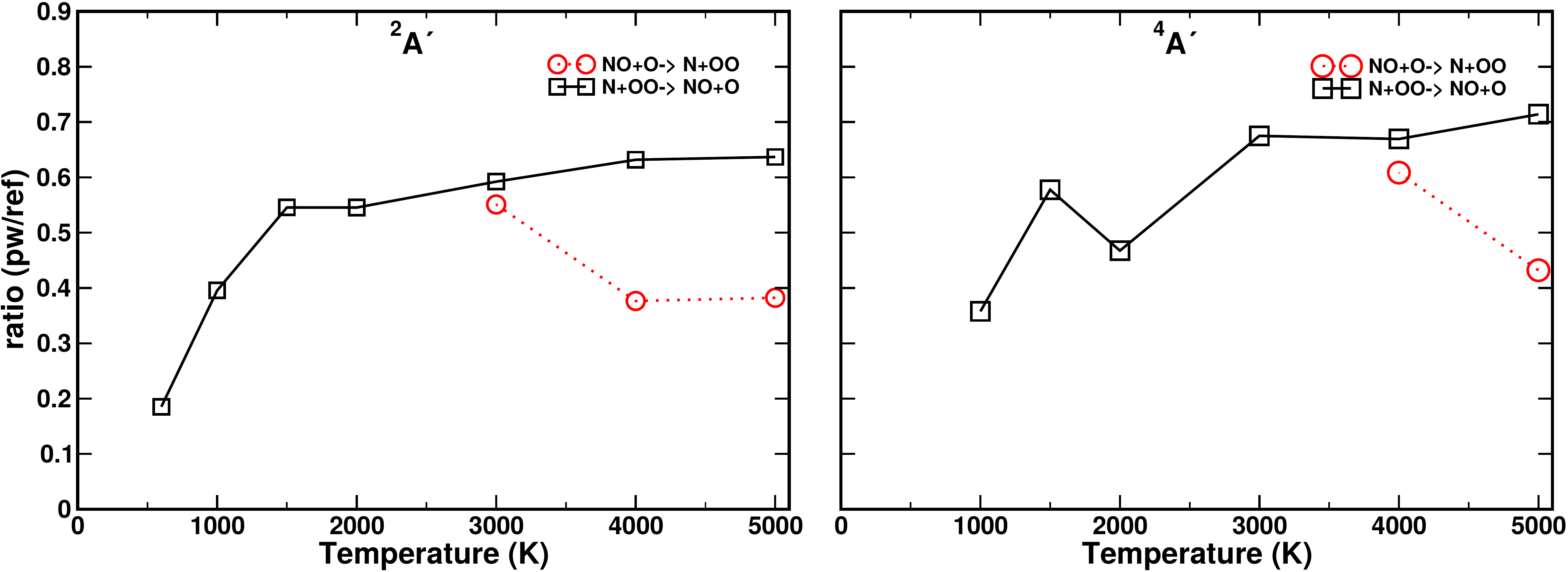}
\caption{Ratio between the present QCT and previously calculated
  ICVT\cite{Says2002} thermal rates as a function of temperature.  For
  the $^2$A$'$ (left) and $^{4}$A$'$ (right) using Gaussian binning
  for the forward (N($^4$S) + O$_2$(X$^3 \Sigma^-_g \rightarrow$
  O($^3$P) +NO(X$^2 \Pi$, solid line) and reverse (O($^3$P) + NO(X$^2
  \Pi) \rightarrow$ N($^4$S) +O$_2$(X$^3 \Sigma _g^-$), dotted line)
  reaction.}
\label{fig:sifig8t}
\end{figure}

\begin{figure}[h]
\centering
\includegraphics[scale=0.6]{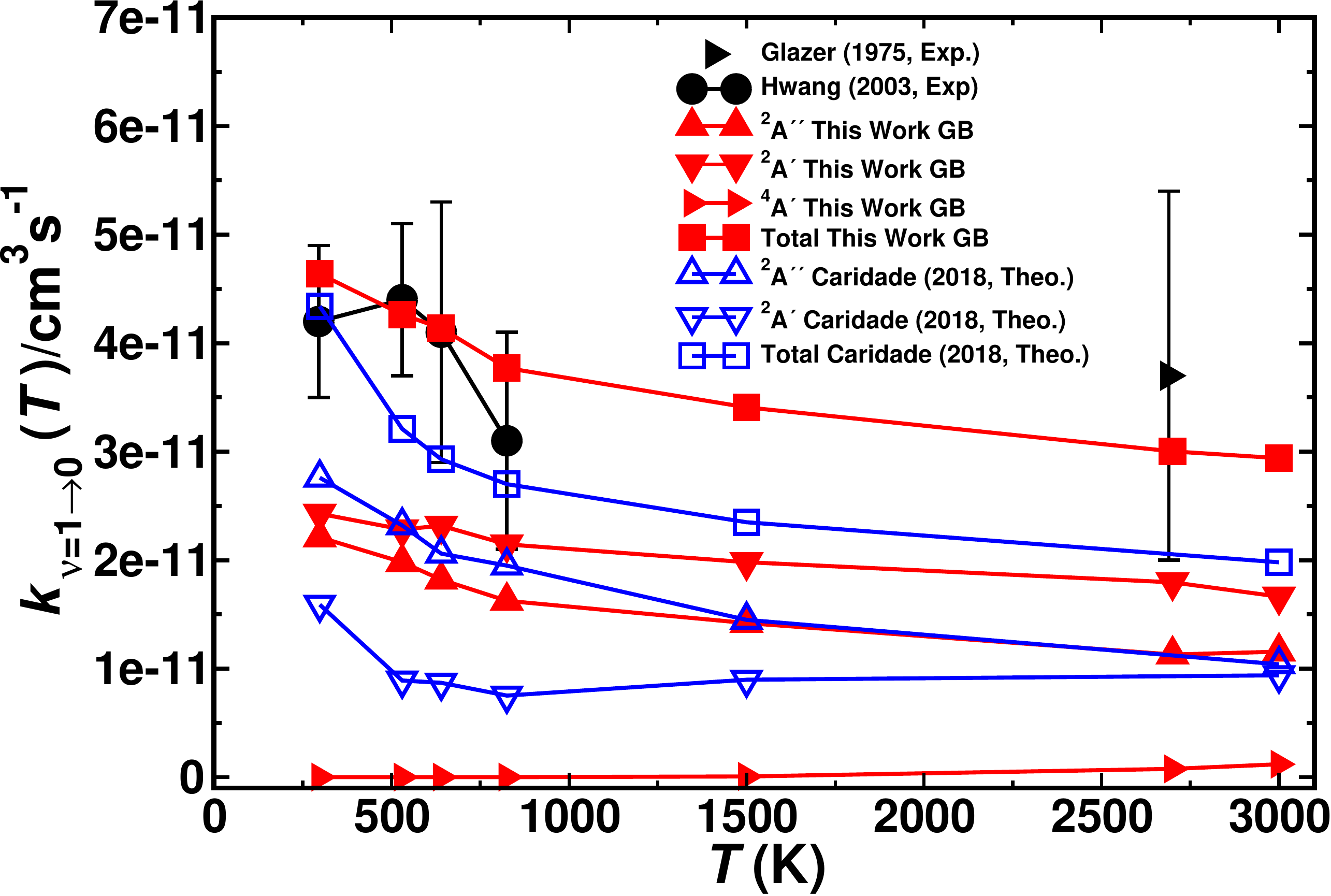}
\caption{Vibrational relaxation rates for O+NO($\nu=1$) $\rightarrow$
  O+NO($\nu'=0$) computed on $^2$A$'$ (triangle down), $^2$A$''$
  (triangle up) and $^4$A$'$ (triangle right) PESs. Results from this
  work are shown as filled symbols while from Ref.\cite{Caridade2018}
  are shown as open symbols. The black filled symbols with error bars
  show the experimental results.}
\label{fig:sifig10t}
\end{figure}

\begin{figure}
\centering
\includegraphics[scale=0.7]{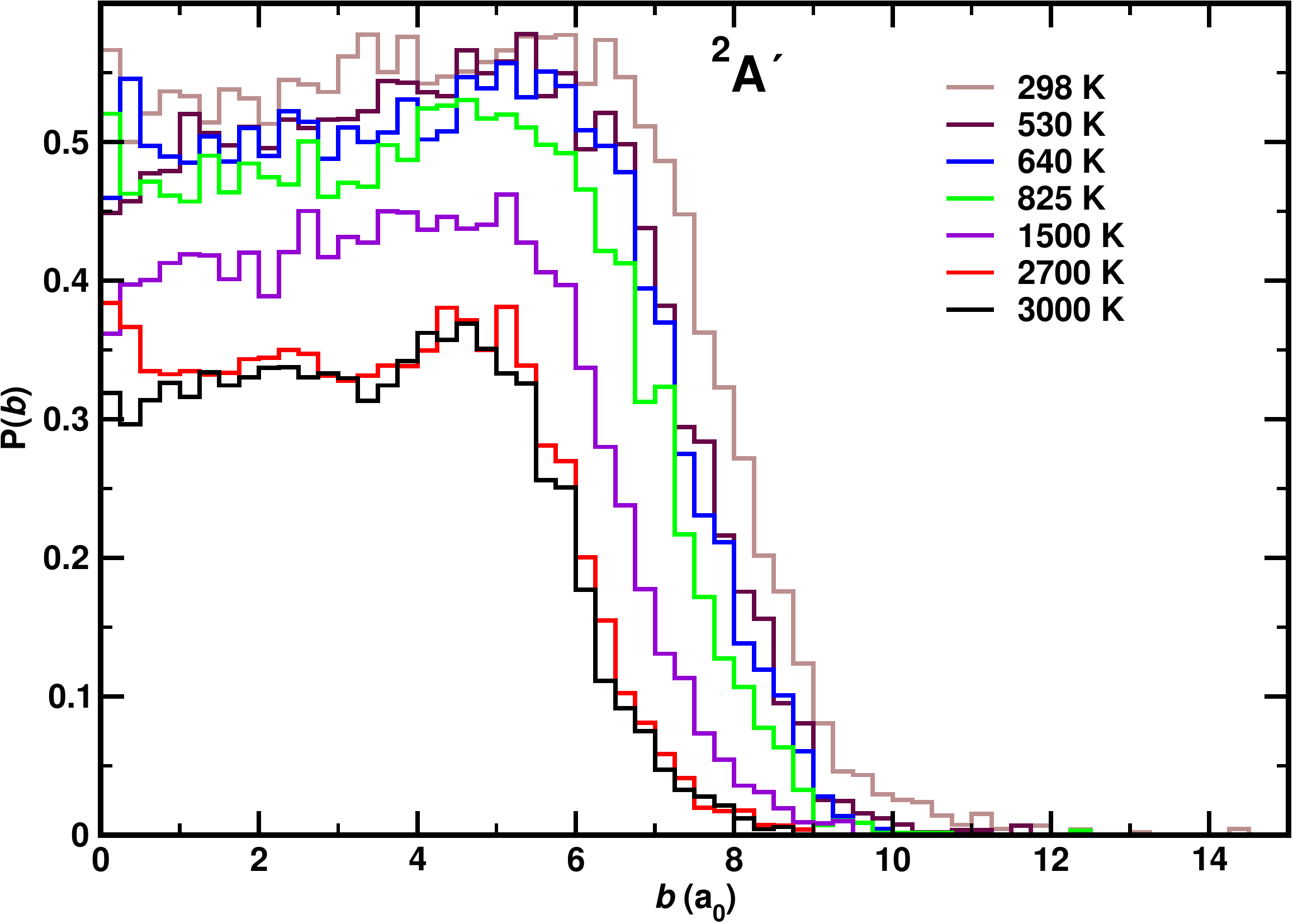}
\caption{Opacity function (probability of vibrational relaxation as a
  function of impact parameter) for the relaxing trajectories (O +
  NO($\nu=1$) $\rightarrow$ O + NO($\nu = 0$)), with or without oxygen
  atom exchange on the $^2$A$'$ PES at different temperatures. With
  increasing $T$ the probability decreases and $b_{\rm max}$ - the
  value for which vibrational relaxation still occurs - shifts to
  smaller values of $b$. The opacity function $P(b)$ indicates that at
  300 K $\sim 50$ \% of all trajectories starting in O + NO($\nu=1$)
  relax to O + NO($\nu = 0$) for small values of $b$ whereas at 3000 K
  only 30 \% relax.}
\label{fig:sifig9t}
\end{figure}

\begin{figure}
\centering \includegraphics[scale=0.8]{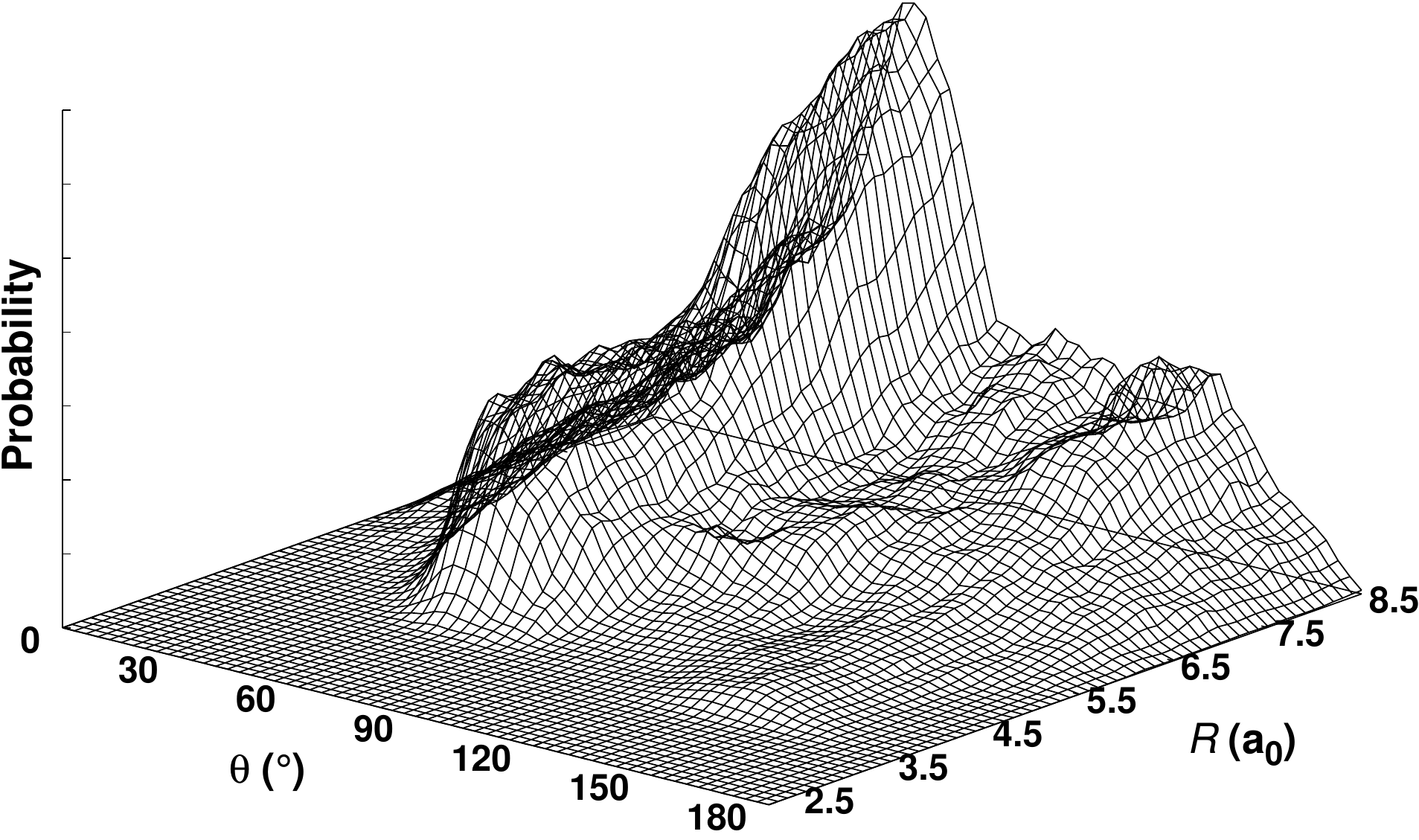}
\caption{Distribution of the vibrationally nonrelaxing trajectories in
  $(R,\theta)$, i.e.  O + NO($\nu=1$) $\rightarrow$ O + NO($\nu\neq
  0$), with or without oxygen atom exchange and N+O$_2$. The distance
  $R$ is the oxygen atom-to-NO(center of mass) distance. Probability
  densities are calculated for all nonrelaxing trajectories and only
  up to the time satisfying the criterion that the sum of the three
  inter-nuclear distances is less than 9.5 a$_0$.}
\label{fig:sifig9}
\end{figure}

\begin{figure}
\centering
\includegraphics[scale=0.8]{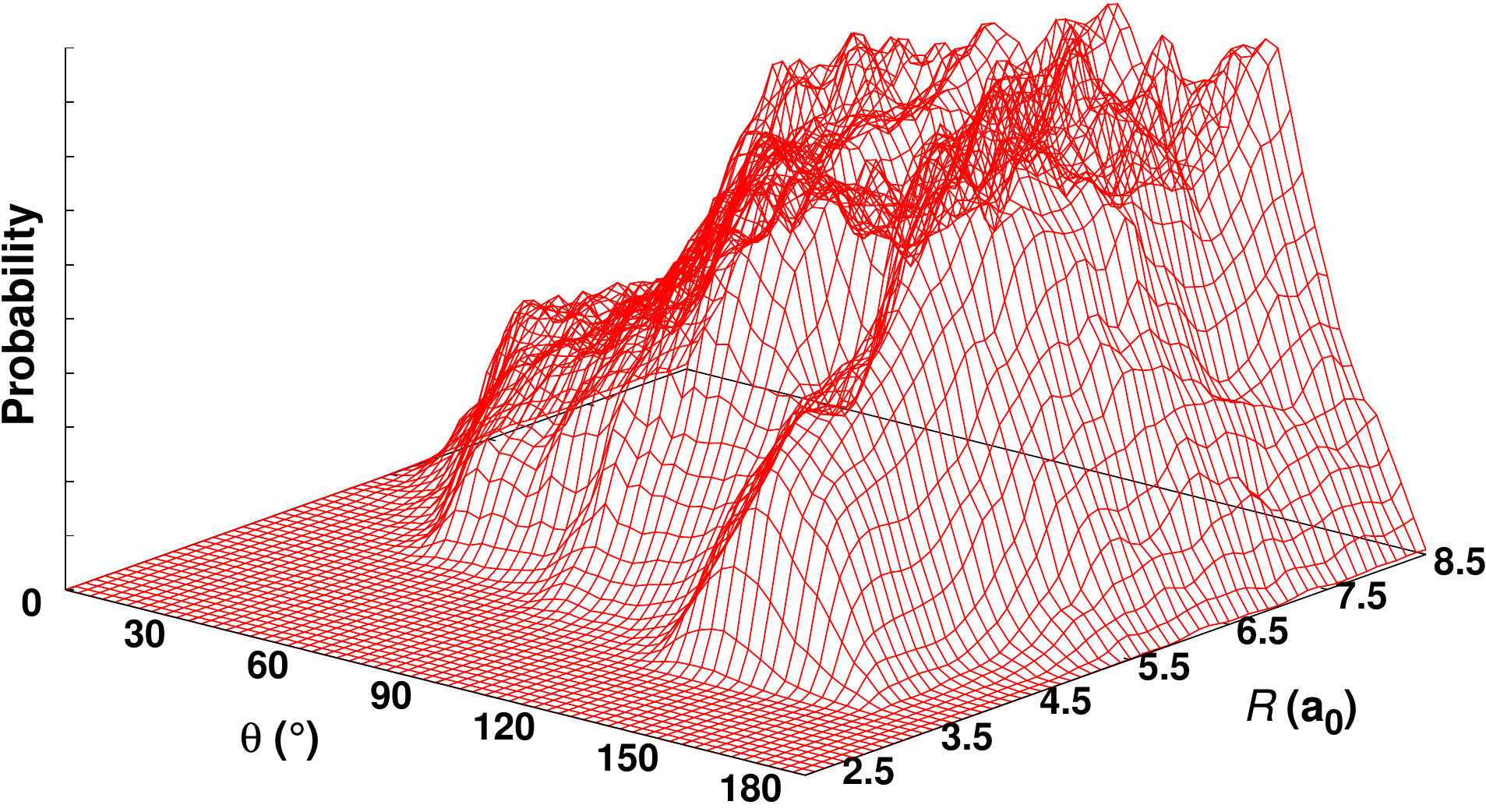}
 \caption{Distribution of the vibrationally relaxing trajectories in
   $(R,\theta)$, i.e.  O + NO($\nu=1$) $\rightarrow$ O + NO($\nu=0$),
   with or without oxygen atom exchange. The distance $R$ is the
   oxygen atom-to-NO(center of mass) distance. Probability densities
   are calculated for all relaxing trajectories and only up to the
   time satisfying the criterion that the sum of the three
   inter-nuclear distances is less than 9.5 a$_0$.}
\label{fig:sifig10}
\end{figure}

\begin{figure}
\centering
\includegraphics[scale=1.1]{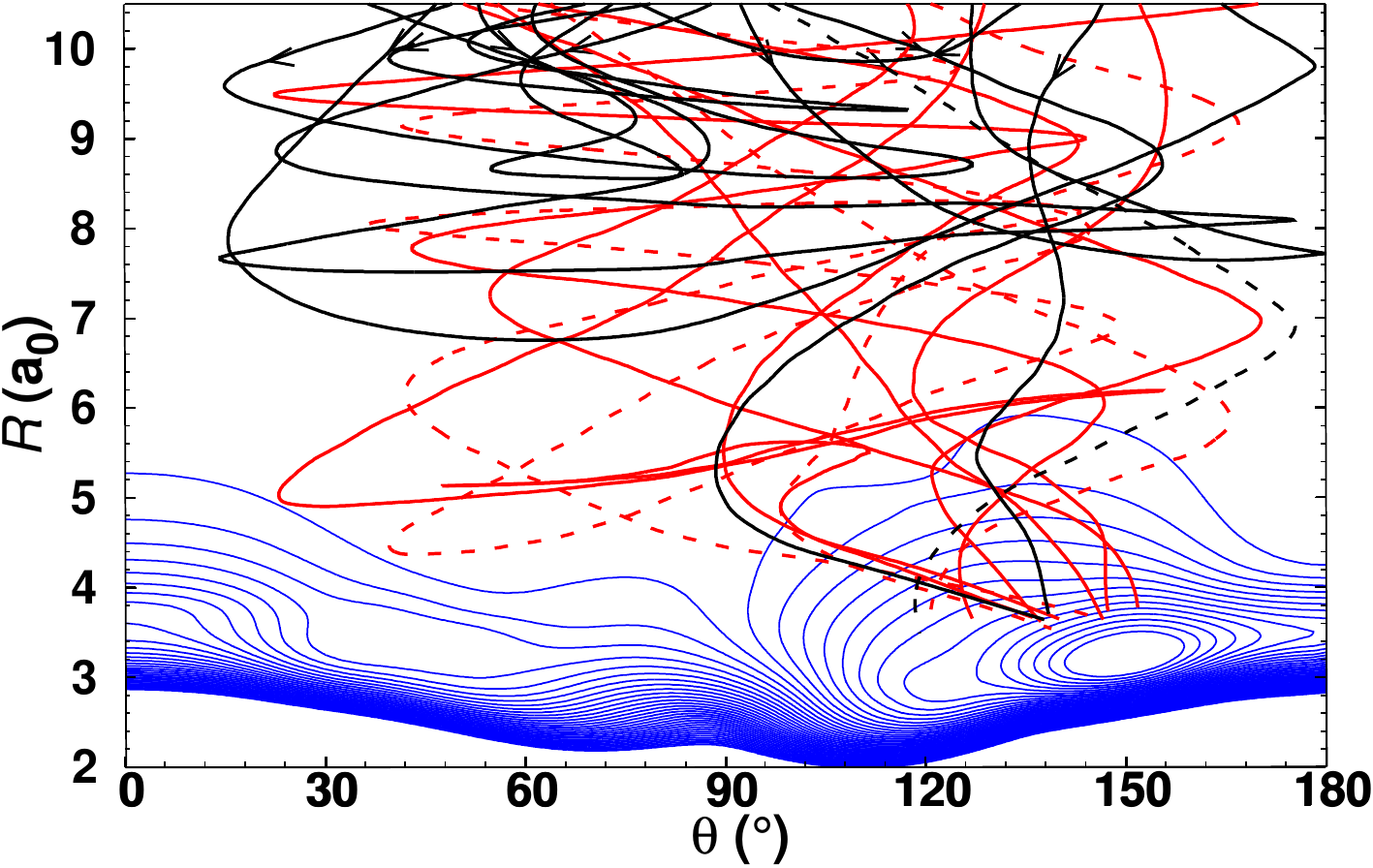}
\caption{Projection of vibrationally relaxing (red) and nonrelaxing
  (black) O+NO collision trajectories onto the $^2$A$'$ PES (blue
  isocontours) as a function of $(R, \theta)$. For each $(R,\theta)$
  combination the energy of the structure with lowest energy for $r
  \in [2.03,2.39]$ (covers the classical turning points of the $v_{\rm
    NO} = 1$ vibration, which are at $r_{\rm min}=2.046$ a$_0$ and
  $r_{\rm max}=2.370$ a$_0$) is used. Ten random trajectories are
  shown for each of the cases. Trajectories are shown only up to the
  time satisfying the geometrical criterion that the sum of the three
  inter-nuclear distances is less than 9.5 a$_0$. Reactive (oxygen
  exchange or O$_2$ formation) trajectories are shown as dashed
  lines.}
\label{fig:sifig11}
\end{figure}

\begin{figure}
\centering
\includegraphics[scale=1.1]{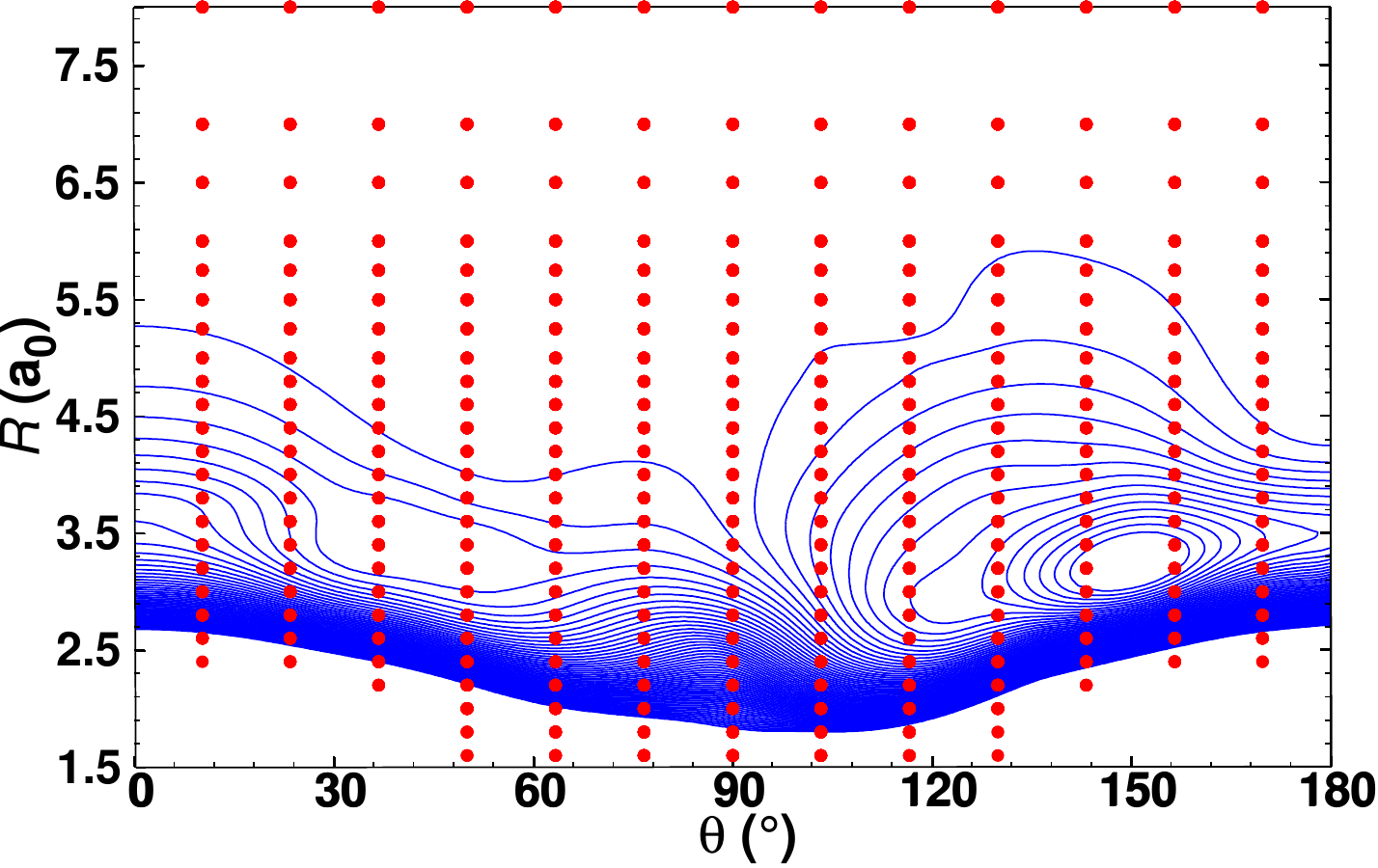}
\caption{Contour diagram of the relaxed PES (as a function of $R$,
  $\theta$ and $r = 2.03 - 2.39$ a$_0$) of the $^2$A$'$ PES for the
  O+NO channel.  The {\it ab initio} grid points used in constructing
  the RKHS are shown as red filled circles at which MRCI+Q
  calculations were carried out.}
\label {fig:sifig12}
\end{figure}

\pagebreak
\bibliography{refs}